\documentclass[11pt]{article}
\usepackage[a4paper, left=2cm, right=2cm, top=2cm, bottom=2cm]{geometry}
\usepackage{epsfig}
\usepackage{cite}
\usepackage{xcolor}
\usepackage{colortbl}
\usepackage{amsfonts}
\usepackage{float}
\usepackage{amsmath,amsthm,amssymb}
\usepackage{multicol}
\usepackage{enumerate}
\usepackage{sidecap}
\usepackage{graphicx}
\usepackage{tikz}
\usepackage{array}
\usepackage{tabularx}
\usepackage{calc}
\usepackage{caption}
\usepackage{subcaption}
\usepackage{cleveref}

\usepackage[scaled]{helvet}
 
\usepackage[T1]{fontenc}

\title{Data-Based Dynamical Systems Reconstruction: An Adequacy/Reliability Test}

\author{
Guillermo Capobianco\\
Instituto de Matem\'atica (INMABB), Departamento de Matem\'atica,\\
Universidad Nacional del Sur (UNS) -- CONICET\\
Alem 1253, Bah\'ia Blanca 8000, Buenos Aires, Argentina
\and
Ulises Chialva\\
Departamento de Matem\'atica, Universidad Nacional del Sur (UNS)\\
Alem 1253, Bah\'ia Blanca 8000, Buenos Aires, Argentina
\and
Horacio G. Rotstein\\
Federated Department of Biological Sciences\\
New Jersey Institute of Technology \& Rutgers University\\
Newark, NJ 07307, USA
}

\date{June 2026}

\begin{document}
\maketitle

\begin{abstract}In this work, we address the problem of validating the reconstruction of a stochastic system from noisy data. We demonstrate the limitations of criteria based solely on the loss function or on standard metrics used for reconstructing deterministic dynamics. We also propose an exploratory approach, based on a two-step test, which allows for a general assessment of the reconstruction without relying on arbitrary error-tolerance thresholds. However, we discuss how system degeneracy and non-identifiability, together with features intrinsic to stochastic dynamics, impose certain constraints on the application of this test.
\end{abstract}

\section{Introduction}
The problem of reconstructing dynamical systems from a limited set of experimental orbits has recently gained significant attention driven by the emergence of new techniques such as artificial neural networks, machine learning, improved neural network training routines, and continuous systems approaches  such as Koopman operators \cite{song2016training,durstewitz2017state,koppe2019identifying, durstewitz2023reconstructing,brenner2023multimodal,kramer2021reconstructing,zhangtangcorpetti,songyangwang16,brunton2021modern}.
Dynamical systems reconstruction aims at revealing the ``rules" that govern the generation of certain empirical (target) time series, and therefore build data-driven mathematical and computational models.
It provides a framework for the  identification of  attractors and other topological and geometric properties of the system's phase space that govern its long-term behavior as well as the sensitivity to external perturbations and  parameter variations, thus offering a deeper understanding of how predictions relate to dynamic mechanisms than approaches limited to the replication of the target data. While the goal of dynamical systems reconstruction goes beyond data replication and prediction, the process is often referred to as time series reconstruction. 

Regardless of the  data fitting tools,  whether  using a discrete (typically a neural network) or continuous (e.g., polynomials) model, the dynamical systems reconstruction process involves the minimization of a loss function representing a distance, or discrepancy between the target and the predicted data \cite{durstewitz2023reconstructing,Goodfellow-et-al-2016}. Loss functions are typically constructed ad hoc,  are adapted to the data and problem at hand, and use a set of attributes (e.g., mean, variance, frequency, the full data set) that are considered to be appropriate to characterize the target data. 
 
 Under rather general conditions, the reconstruction of   deterministic trajectories (typically noisy, but assumed to be generated by deterministic systems) in the phase-space is theoretically guaranteed \cite{takens2006detecting,cybenko1989approximation}, but there is no similar theory when the generating system cannot be safely assumed to be deterministic. In the absence of a theory to fall back to, the validity of reconstructions need to be assessed on a case by case basis and validation becomes an integral part of the reconstruction procedure \cite{durstewitz2023reconstructing}.

There are generally two strategies to approach this. 
In one approach,  no additional validation of the reconstruction is pursued outside the loss function minimization (confidence on the loss function minimization is considered to be the validation of its own result). The other approach  consists of validating the reconstruction independently of the training process by using data that has been excluded from the training set.  This is typically achieved  by using metrics that are computationally more expensive, but more precise than these used during training such as the fractal dimension of the data, the Kullback-Leibler divergence, the Hellinger distance and the Wasserstein distance \cite{deza2009encyclopedia}. The latter three are evaluated using histograms of the data set \cite{koppe2019identifying, hemmer2024optimal}. 

Two often overlooked major challenges arise in both strategies: degeneracy and  determination of the reconstruction quality. The first one refers to the ability of different models to generate the patterns having the same attribute values and therefore fit the target data equally well \cite{bellman1970structural, cobelli1980parameter, prinz2004similar, raue2009structural, lederman2022parameter}. The second one results from the lack of universal criteria to determine what  constitutes a good (or bad) reconstruction. Because different metrics focus on different sets of attributes and differentially weight them, their use can provide different answers to the reconstruction quality question. Even if one model results to be a better reconstruction than another according to a selected metric, this may not hold for another metric.

What constitutes an adequate reconstruction is a central question with broader implications.  If we formalize the answer by stating that a small error for a given metric is enough to guarantee an adequate reconstruction, then the question arises of what makes an error small and what this  small error tells us about the ability of the putative reconstructed system to generate the data.
Typically, the acceptable error threshold varies depending on the context and the putatitve system's dynamic structure. A certain margin of error may be acceptable for some systems (e.g., chaotic), but unacceptable for others (e.g., linear, with a single stable fixed-point). Different metrics may produce different levels of error for the same data. 
Moreover, the error levels for degenerate systems are expected to be comparable. Because of this lack of precision and the presence of ambiguity, making universal comparisons between reconstruction methods becomes very difficult, if not virtually impossible. Note that  we use the term ``adequate" rather than ``accurate" to acknowledge this issue. 

The goal of this paper is to address these issues.  We develop a two-stage test that quantifies how well the candidate reconstructed system reproduces the target data. This adequacy/reliability (AR) test combines the explorative Tukey interquartilic range method \cite{tukey1977exploratory} and the statistical sign test \cite{conover1999practical} to assess  how well the original data ``blends in" with a dataset generated by the candidate system. More specifically, the AR test  determines how well both the trajectory coverage and recurrence in the phase-space match, and the statistical significance of the deviations  between the experimental data and the dynamics simulated by using the reconstructed system. This test is universal in the sense that it is agnostic to metrics (e.g., distances, divergences) and therefore some of the issues highlighted above are avoided, and it can be combined with additional metrics to resolve ambiguities when they arise. 

\label{test_summary}

The outline of this paper is as follows.
First, we analyze the problem of reconstructing noisy dynamics that preserve the \emph{geometric structure} of the experimental data, and we show that the usual metrics are insufficient due to the intrinsic variability of the putative generating systems. Then, we introduce the two-stage AR test mentioned above to  validate reconstructions from noisy data. Subsequently, we discuss an example consisting of the reconstruction of a chaotic system supplemented with noise validated by the two-stage AR test, and we analyze the robustness of the results. After that, we show how the degeneracy of certain systems constrains the applicability of the test in some cases and discuss possible alternatives. Finally, we discuss the results and their implications.

\section{Methods}

\subsection{Models}
\label{models}

We use four representative models to generate ground truth data and analyze the validity of a putative dynamical systems reconstruction using the AR test. The Chua \cite{chua1986double,matsumoto2003chaotic} and Lorenz  \cite{lorenz1963,strogatz2024nonlinear} models are classical examples of three-dimensional dynamical systems that exhibit chaotic behavior for appropriate parameter choices. The FitzHugh-Nagumo (FHN) \cite{fitzhugh1960thresholds,sherwood2022fitzhugh} and  Lambda-Omega ($\Lambda\Omega$) \cite{glass1988clocks} models are prototypical two-dimensional nonlinear sustained oscillators. The FHN model has been used in many fields of science, in particular as a phenomenological simplification of the Hodgkin-Huxley model \cite{izhikevich2007dynamical,ermentrout2010mathematical} for neuronal dynamics, and can display oscillatory patterns that range from sinusoidal-like to relaxation type oscillations. The $\Lambda\Omega$ model we use is  a generalization of the normal form of the supercritical Hopf-bifurcation that has been used as a prototypical model for sustained oscillations and provides a canonical framework for the study of degeneracy in oscillatory systems \cite{lederman2022parameter}. In the literature, it is also referred to as the Stuart-Landau oscillator \cite{landau1944,stuart1958}.

In the four models below, \( x \), \( y \) and \( z \) are the state variables, \( t \) is time and the ``dot" indicates derivatives with respect to \( t \).

\bigskip
\noindent
\textbf{Chua model:} In this work, we use a smoothed (cubic-order) version of the Chua circuit, following \cite{zhong1994implementation,pivka1996lorenz}:

\begin{equation}
\begin{aligned}
\dot{x} &= \alpha\left(y-m_0x-m_1x^3\right),\\
\dot{y} &= x-y+z,\\
\dot{z} &= -\beta y-hz,
\end{aligned}
\end{equation}

\noindent
where \( \alpha \), \( m_0\), \(m_1 \), \( \beta \) and \( h \) are constants. In general, it is assumed that $m_0 m_1 < 0$  \cite{pivka1996lorenz}.

\bigskip
\noindent
\textbf{Lorenz model:} It is described by
\begin{equation}
\begin{aligned}
\dot{x} &= \sigma(y-x),\\
\dot{y} &= x(\rho-z)-y,\\
\dot{z} &= xy-\beta z,
\end{aligned}
\end{equation}

\noindent
where \( \sigma \), \( \rho \) and \( \beta \) are constants. We use the classical parameter values $\sigma = 10$, $\rho = 28$ and $\beta = \dfrac{8}{3}$ \cite{lorenz1963}.

\bigskip
\noindent
\textbf{FHN model:} It is described by
\begin{equation}
\begin{aligned}
\dot{x} &= x-\frac{x^3}{3}-y+I,\\
\dot{y} &= \frac{x+a-by}{\tau},
\end{aligned}
\label{fhn_eq}
\end{equation}

\noindent
where the parameters \( a \), \( b \), \( \tau \) and \( I \) are constants. In general, $b, \tau > 0$, while $a$ and $I$ are parameters used to determine the location of the equilibrium.

\bigskip
\noindent
\textbf{$\Lambda\Omega$ model:} It is described by
\begin{equation}
\begin{aligned}
\dot{x}\; &= \left(\lambda -br^2\right)x-\left(\omega+ar^2\right)y,\\
\dot{y}\; &= \left(\omega+ar^2\right)x+\left(\lambda -br^2\right)y,\\
r^2 &= x^2+y^2\\
\end{aligned}
\label{lo_eq}
\end{equation}

\noindent
where \( \lambda \), \( \omega \), \( a \) and \( b\) are nonnegative constants. 

\bigskip
\noindent
\textbf{Modified Piecewise Linear Recurrent Neural Network (PLRNN):} We use a modified version of the  PLRNN  introduced in \cite{durstewitz2017state} and subsequently used in \cite{koppe2019identifying, monfared2020transformation}. We consider the original PLRNN  with an additional hidden layer, obtaining a discrete-time system of the form
\begin{equation}
z_{t+1} = A\, z_{t} + W_1\textrm{ReLU}\left(W_2\,z_t + h_2\right) + h_1 + \epsilon
\label{plrnn_net}
\end{equation}

\noindent
where $z_t,h_1\in\mathbb{R}^n$, $A\in\mathbb{R}^{n\times n}$, $W_1\in\mathbb{R}^{n\times m}$, $W_2\in\mathbb{R}^{m\times n}$, $h_2\in\mathbb{R}^m$ and $\epsilon$ is a noisy term.

\bigskip
\noindent
\textbf{Gaussian noise:} A white (Gaussian) noise term of the form  $\sqrt{2D}\, \eta(t) $ where \( D \) is the variance and \( \eta \in {\cal N}(0,1)\) was added to the first equation in all models. \label{noise_protocol}

\subsection{Numerical simulations}
The simulations were performed using the modified Euler-Heun method (a Runge-Kutta method of order 2) \cite{suli2003introduction} for deterministic systems and the stochastic Euler-Heun scheme \cite{honeycutt1992stochastic} for stochastic systems.  The integration step is specified in each case. All routines were written in Python, and the training routines were implemented using the PyTorch framework.


\subsection{The two components of the AR test}

The AR test we develop in this paper  consist of two components, the explorative data analysis Tukey's interquartile range (IQR) test and the statistical sign test.
\label{twotests}
\bigskip

\noindent{\bf Tukey's IQR test:}{ We employ the outlier detection criterion introduced by Tukey \cite{tukey1977exploratory}, based on the interquartile range (IQR). This method relies on the robust characterization of the data dispersion through the first and third quartiles, $Q_1$ and $Q_3$, respectively. Defining the interquartile range as $\mathrm{IQR} = Q_3 - Q_1$, observations are classified as outliers if they fall outside the interval $[Q_1 - f\,\mathrm{IQR},\; Q_3 + f\,\mathrm{IQR}]$, where $f > 0$ is a tolerance parameter controlling the strictness of the criterion. The most common choice is $f = 1.5$, although $k = 3$ is also used to identify more extreme outliers. We use $f=1.5$ in this work.}
\bigskip

\noindent
{\bf Sign test:}{ It is a nonparametric statistical test used to assess whether the median of a distribution differs from a specified reference value \cite{conover1999practical}. Given a sample $\{x_i\}_{i=1}^N$, each observation is classified according to the sign of its deviation from the reference value, assigning a positive or negative sign while discarding zero differences. Under the null hypothesis, the number of positive signs follows a binomial distribution with success probability $p = 0.5$. Deviations from this binomial behavior provide evidence against the null hypothesis, making the test particularly suitable for assessing binomiality in sign-based data without assumptions on the underlying distribution.}
\bigskip

Exploratory (data) analysis refers to the adoption of a flexible, assumption-light approach, centered on data visualization and the use of robust statistical summaries, with the aim of facilitating understanding of the structure of the dataset and guiding subsequent modeling decisions rather than testing specific hypotheses. This is in contrast to  standard statistical treatments, which typically emphasize inference based on probabilistic models and previously established assumptions.

\section{Results}

\subsection{The geometric structure of data: Limitations of the usual metrics for comparison between data set generated by stochastic systems}
\label{geometricstructure}
 
When analyzing the quality of a reconstruction  one seeks to determine whether the range of values of  the reconstructed and target data are ``close enough" and whether the reconstructed and target trajectories ``belong to the same cloud" in the region of the phase-space  where one has observational access (typically one observable variable). Replicating the target data set pointwise is often not an achievable nor even a desirable goal  giving the inherent variability of the data,  the structure of the putative dynamical systems, which can be stochastic or chaotic, or the effects of initial conditions (e.g., out-of-phase oscillations that are otherwise identical). Therefore, one needs to resort to data distributions  for the observable attributes that are considered to be descriptive enough of the target data and compare said distributions between the target and reconstructed data. This is captured by the notion of \emph{geometric structure} \cite{koppe2019identifying,durstewitz2023reconstructing} that  refers to the values taken by a signal (target or reconstructed) and the relative frequency of occurrence of each value. Comparison between the geometric structures of the target and the reconstructed data allows to  asses the quality of the reconstruction for systems that have been properly reconstructed (e.g., have been generated by the same dynamical system) even if the target trajectory cannot be exactly replicated (Fig. \ref{basic_test}).

The typical approach consists of comparing the histograms (Fig. \ref{basic_test}-B) for the target and reconstructed data (Fig. \ref{basic_test}-A).
This requires the use of a metric that is appropriate for the type of distributions at hand, which may vary in shape, but still represent outcomes of the same dynamical system. Although this method has been widely used to analyze the reconstruction of deterministic systems\cite{koppe2019identifying} and it benefits from ergodicity for some systems for which the geometric and temporal structures are related (e.g., some chaotic systems), it has limitations when applied to noisy target signals. In stochastic models, trajectory uniqueness does not hold, which renders pointwise comparison methods (such as those based on mean squared error loss functions) inapplicable, even for simple systems such as a noisy oscillator. Furthermore, the presence of noise induces unavoidable fluctuations in the distributions obtained from orbit histograms, making it difficult to employ metrics whose validity relies on attaining zero or arbitrarily small values.

\begin{table}[H]
\centering
\begin{tabular}{lll}
{\bf A} & {\bf B} & {\bf C} \\
\includegraphics[width=0.32\textwidth]{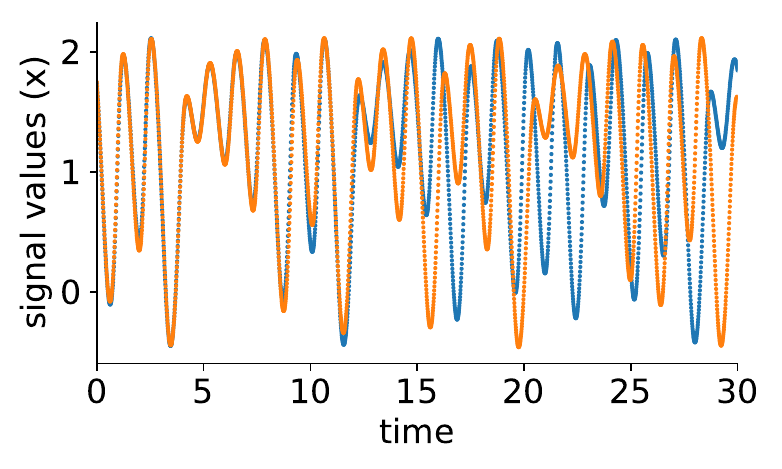} &
\includegraphics[width=0.32\textwidth]{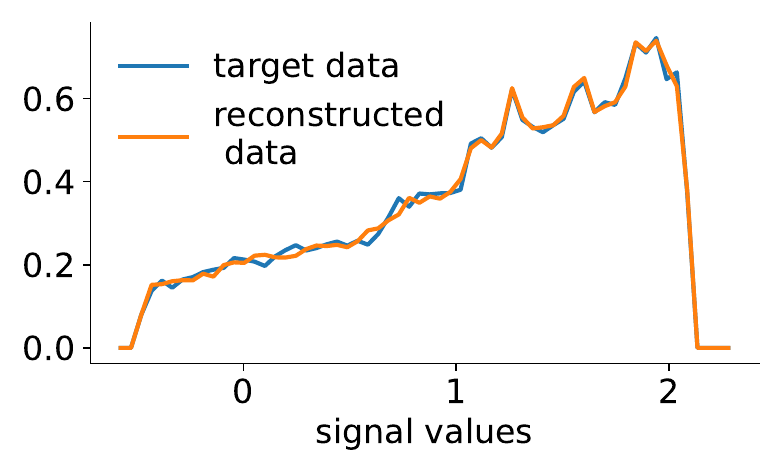} &
\includegraphics[width=0.32\textwidth]{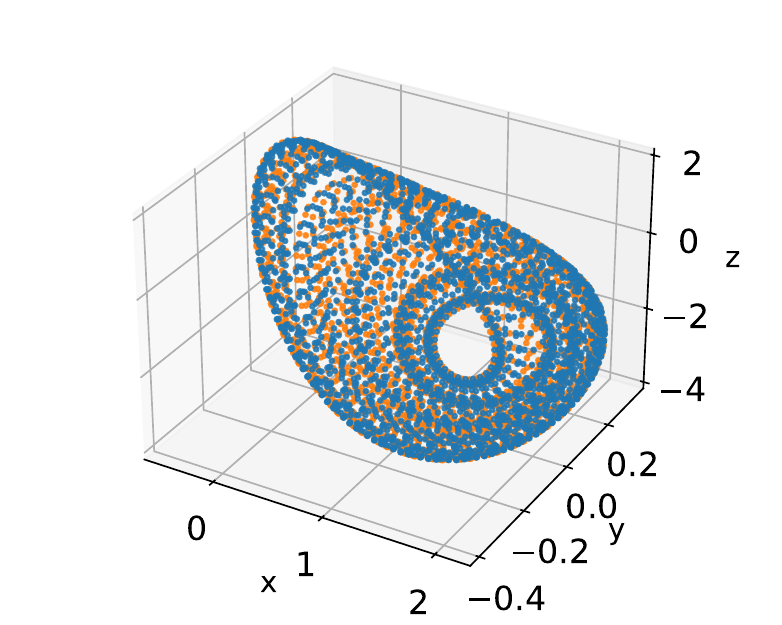} 
\\
\end{tabular}
\caption{\small{\textbf{The geometric structure of signals: Target and reconstructed data.}  
\textbf{A.} Target (blue) and reconstructed (orange) signals for the Chua model, with $\alpha = 15$, $\beta = 33$, $m_0=-\frac{1}{6}$, $m_1=\frac{1}{16}$ and $h= -\frac{1}{7}$. \textbf{B.} Histograms for the corresponding data in panel A:  range of values covered by the trajectory  (abscisa) and the recurrence of each value (height). \textbf{C.} Phase-space diagram for the two signals. Due to  sensitivity to initial conditions the two signals are significantly different, but the histograms are similar, consistently with the trajectories belonging to the same 'cloud of points' in the phase-space diagram.}}
\label{basic_test}
\end{table}

To illustrate the latter, we analyze two representative models in the presence of noise and three metrics for each model (Fig. \ref{variability_problem}). The models are the FHN oscillator (top row) and the  Lorenz (chaotic) system (bottom row), and  the metrics  are the Kullback-Leibler divergence (left column), the Wasserstein distance (middle column) and the Hellinger distance (right column). The Kullback-Leibler divergence measures information loss between histograms, the Wasserstein distance captures geometric differences between distributions, and the Hellinger distance quantifies differences in values, with the advantage of being bounded in the [0,1] interval \cite{deza2009encyclopedia}.  

For each system, we simulated a test orbit (a signal mimicking the target data) and several additional trials (all starting from the same initial condition as the test orbit, mimicking putative reconstructions). We then computed the three metrics for the histograms of the test orbit and each one of the trial orbits. 
We used the mean and the coefficient of variation $CV$ (the quotient of the standard deviation and the mean)  to quantify the variability across trials and analyzed the robustness of each one of the metrics after removing outliers that are not representative of the behavior and may disrupts the results by using Tukey's test \cite{tukey1977exploratory}.

As expected, the Lorenz system  exhibits higher variability than the FHN system (Fig. \ref{variability_problem}). However, in both cases, the three metrics  fail to robustly recognize that all  the trial time series  were generated by the same system.  For the FHN system, the three metrics show a moderate $CV$, around $20-30\%$, while for the Lorenz system, the three metrics show higher $CV$ .  In this case, divergence values close to $0$ coexist with others that double the mean value, and belong to the typical range, so they cannot be discarded as exceptional cases. In the case of the FHN system, the metrics are more stable but never close to 0 (the lower bound of their typical range is greater than $0$).

\begin{table}[H]
\centering
\begin{tabular}{lll} 
{\bf A1} & {\bf A2} & {\bf A3} \\
\includegraphics[width=0.3\textwidth]{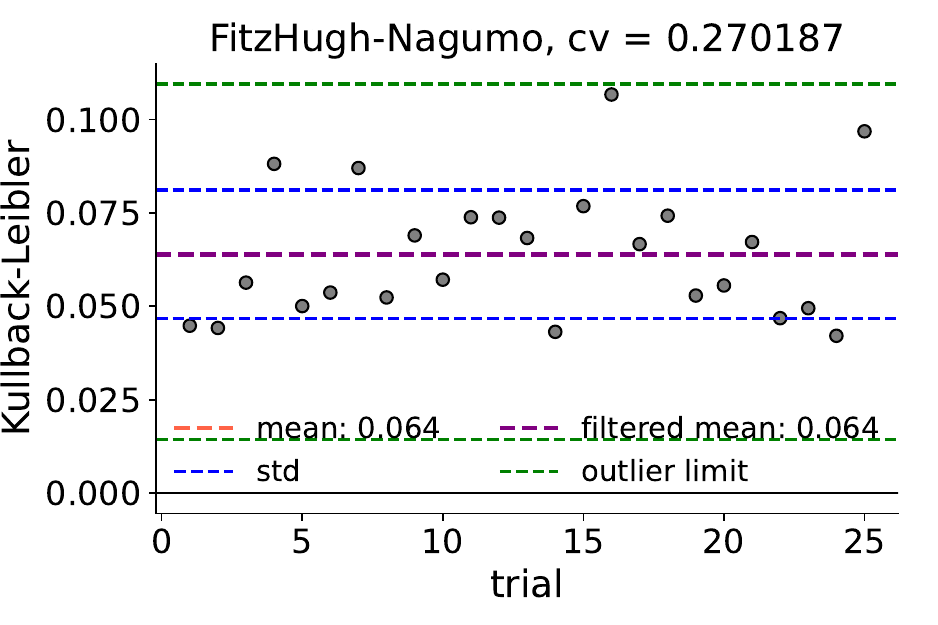} &
\includegraphics[width=0.3\textwidth]{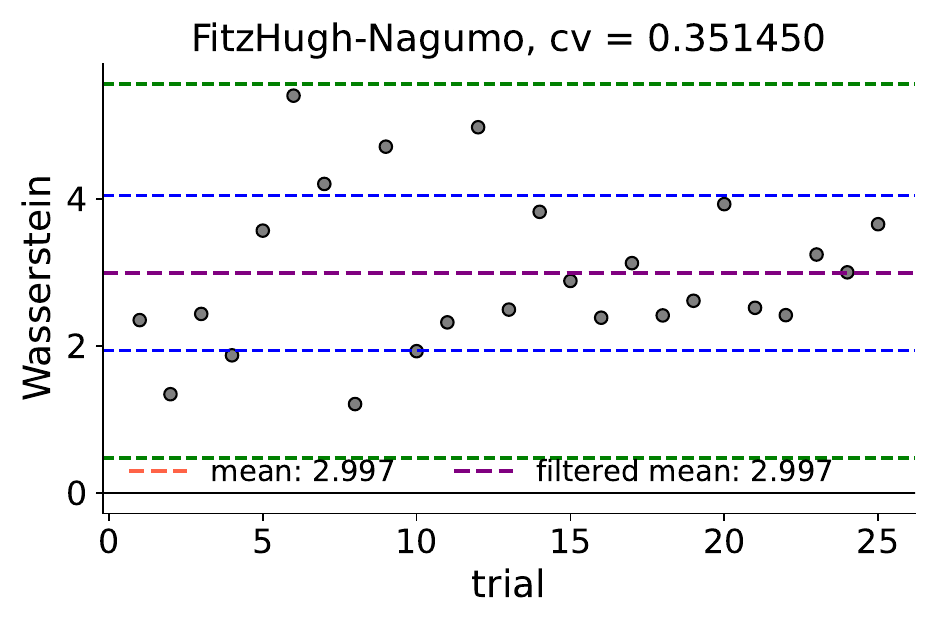}&
\includegraphics[width=0.3\textwidth]{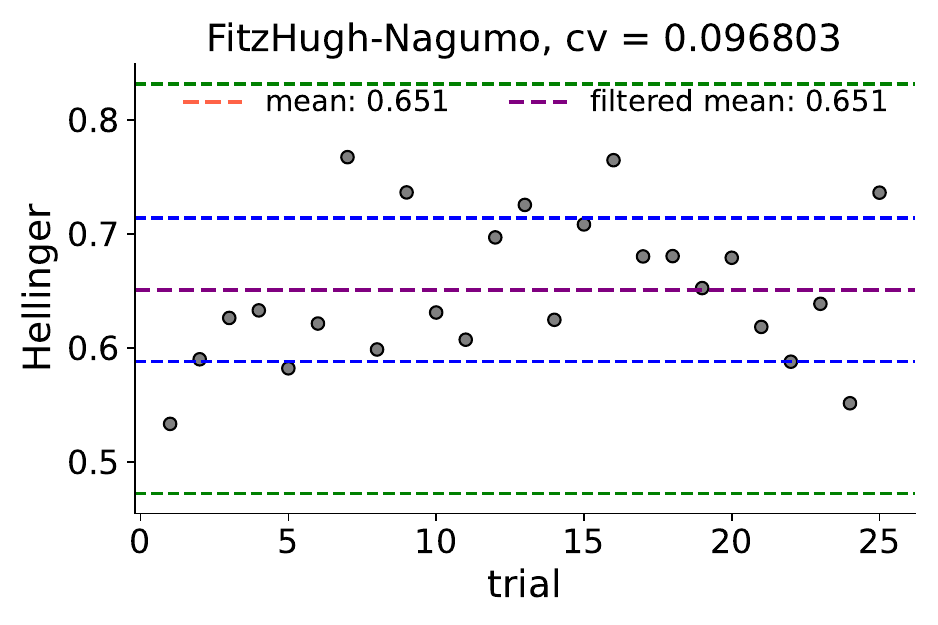} \\
{\bf B1} & {\bf B2} & {\bf B3} \\
\includegraphics[width=0.3\textwidth]{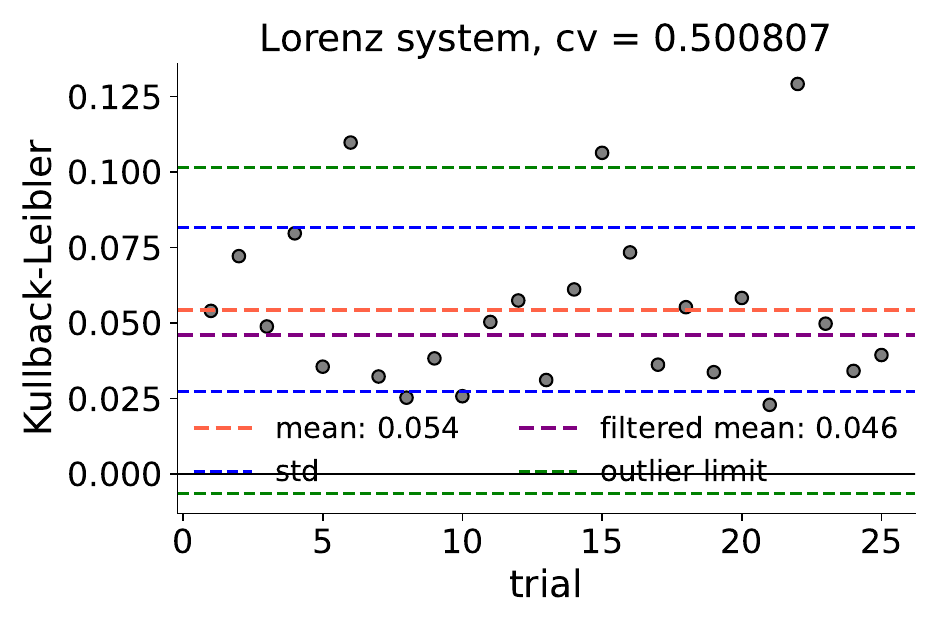} &
\includegraphics[width=0.3\textwidth]{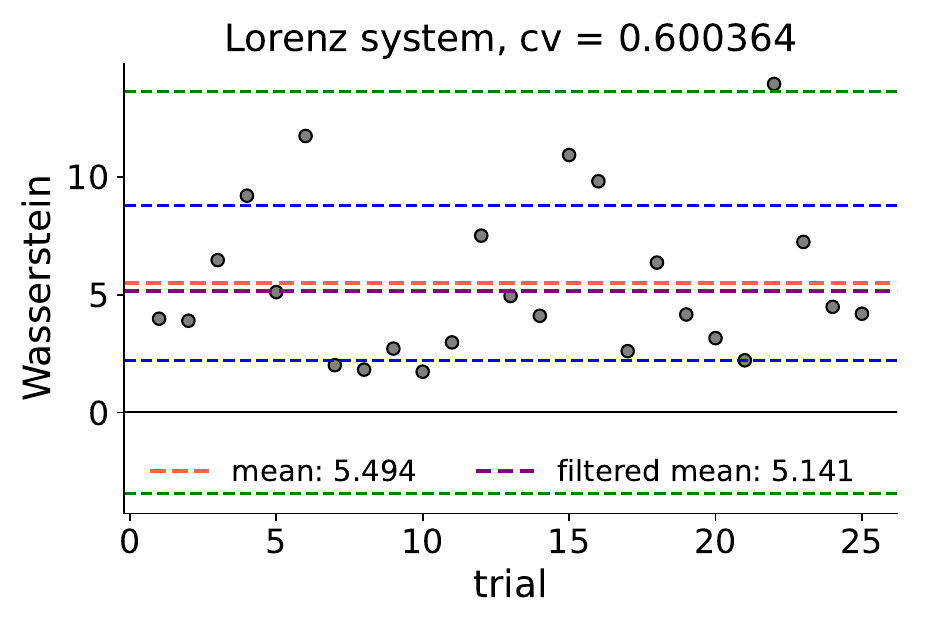}&
\includegraphics[width=0.3\textwidth]{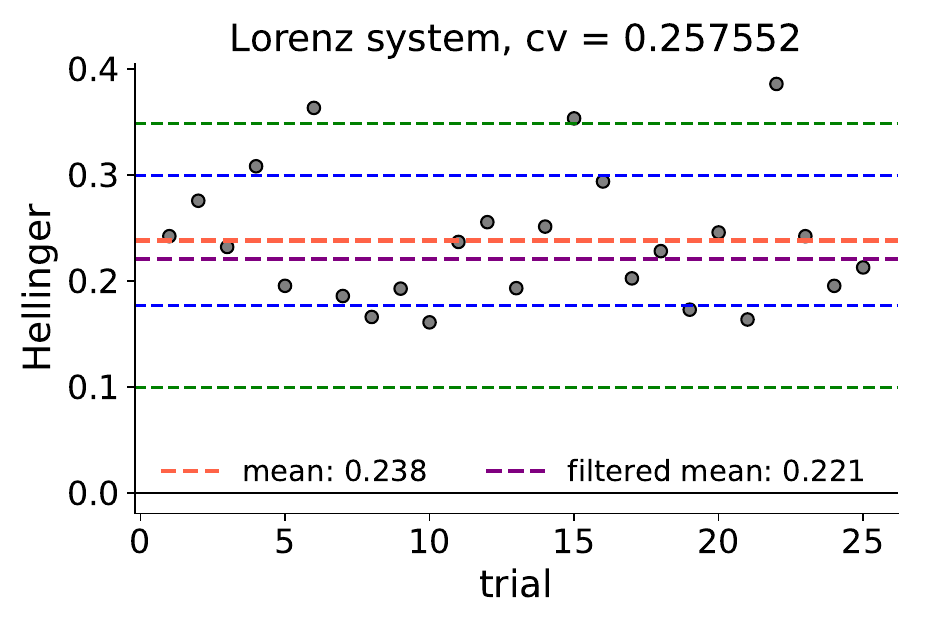} \\
\end{tabular}
\caption{\small {\bf Signal variability affects the robustness of the metrics: Representative models in the presence of noise.} 
For each model, we simulated a test orbit (mimicking the target data) and twenty five additional trials (all starting from the same initial condition as the test orbit, mimicking putative reconstructions). We applied three different metrics to the histograms of the test orbit and each one of the trials. 
We use the mean and the coefficient of variation $CV$ to quantify the variability across trials and analyze the robustness of each one of the metrics after removing outliers that are not representative of the behavior and may disrupts the results by using  Tukey's test. In both models we added white Gaussian noise to the  equation for the variable \( x \).
Representative examples are shown in Fig. \ref{trials_ex}.
{\bf A.} FitzHugh-Nagumo model in the presence of noise. We used the following parameter values: $a = 0$, $b=0.8$, $I=0$, $\tau = 12$, and \( \Delta t = 0.025 \).  For the Gaussian noise, we used mean= 0 and variance = 0.5.
{\bf B.} Lorenz system in the presence of noise. We used the following parameter values: $\sigma = 10$, $\rho = 28$, $\beta = \frac{8}{3}$ and \( \Delta t = 0.005 \). For the Gaussian noise, we used mean = 0 and variance = 50.
{\bf Left column.}  Kullback-Leibler divergence.
{\bf Middle column.} Wasserstein distance.
{\bf Right column.} Hellinger distance.
 }%
\label{variability_problem}
\end{table}

\begin{table}[H]
\centering
\begin{tabular}{c c}
{\bf A} & \includegraphics[width=0.8\textwidth]{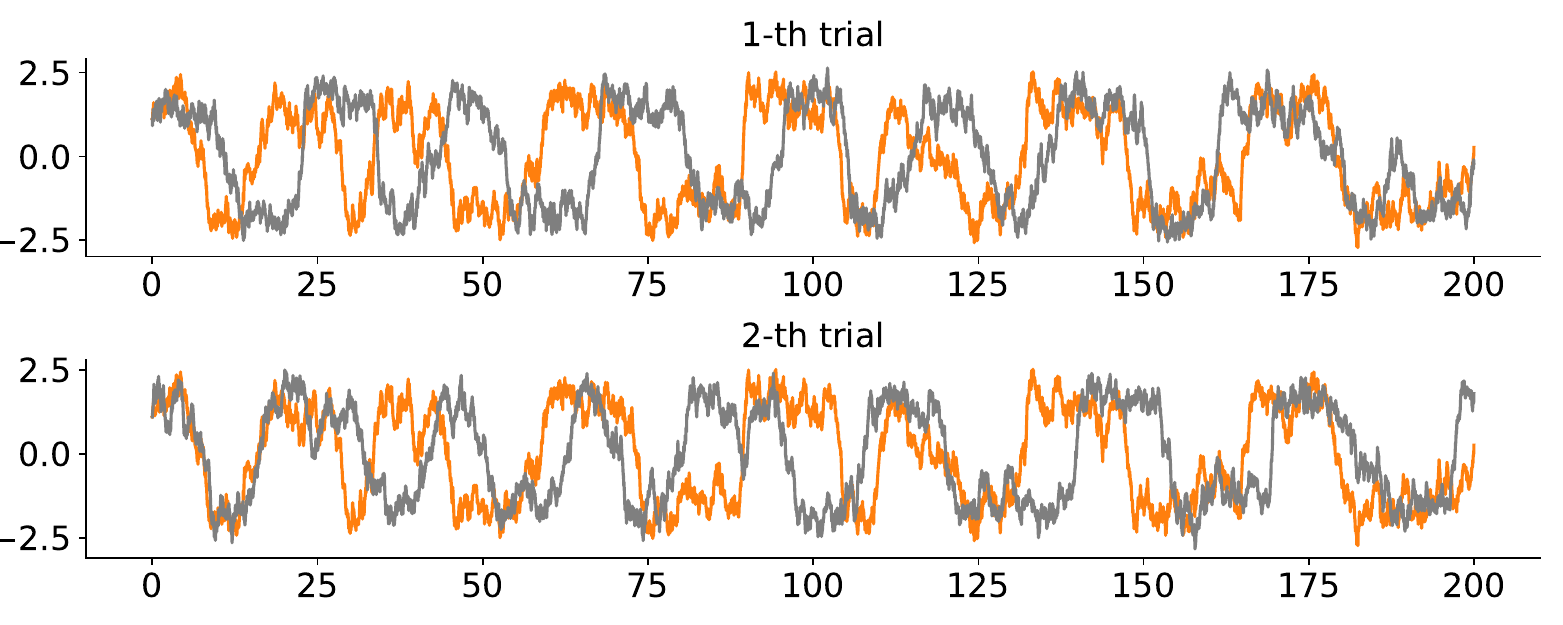} \\
{\bf B} & \includegraphics[width=0.8\textwidth]{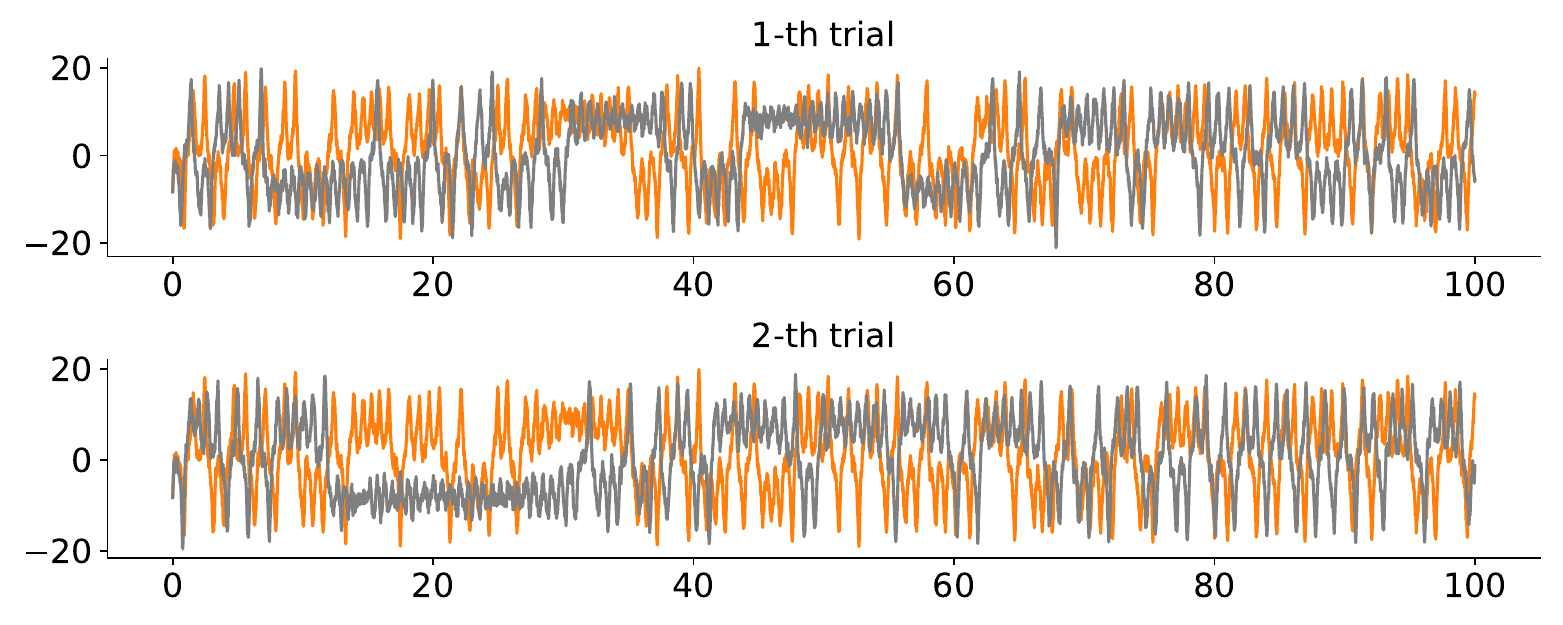} \\
\end{tabular}
\caption{\small {\bf Representative examples corresponding to Fig. \ref{variability_problem}.} The orange curves correspond to the test orbits and the gray curve correspond to the chosen representative trials. 
{\bf A.} FitzHugh-Nagumo model in the presence of noise. 
{\bf B.} Lorenz system in the presence of noise.  
}
\label{trials_ex}
\end{table}

\subsection{The adequacy/reliability (AR) test}

\subsubsection{Statement of the problem}

There is an unknown dynamical system \( S \) for which we only have access to discrete values of a fragment of the (target) time series \( s(t) \) (\( 0 < t < T \)) for an observable variable. 
We aim to analyze the quality of the reconstruction of  \( S \) provided by a model  $M$ by using the time series \( s(t) \) and a set of trials \(x_{trial,n}(t) =   \{x_i(t)\}_{i=1}^{n} \) generated by the model $M$. This putative reconstruction model is assumed to have been obtained by using appropriate methods (e.g., parameter estimation techniques, neural network training).

\subsubsection{Approach}

Our approach consists of analyzing how well the target series \( s(t) \)  ``blends in" within the set of trials 
 \(x_{trial,n}(t)  \).  Specifically, we assess how ``representative" is the histogram \( h_s \) of \( s(t) \) within the family of histograms \( h_{trial,n} = \{h_i\}_{i=1}^n \) for \( x_{trial,n}(t) \). 
Intuitively, we determine whether \( h_s \) is a typical  histogram of a trial generated by the model \( M \) with certain level of confidence. The notion of being ``representative" or ``typical" will be defined more precisely as part of the development of the method. 

All histograms are computed over the same bin range, have the same number of bins (\(K\)) and are normalized such that the total area under each histogram adds up to 1.

\begin{table}[htpb!]
    \centering
    \begin{tabular}{lll}
        \multicolumn{3}{l}{\raisebox{15ex}{\bf A} \includegraphics[scale=0.7]{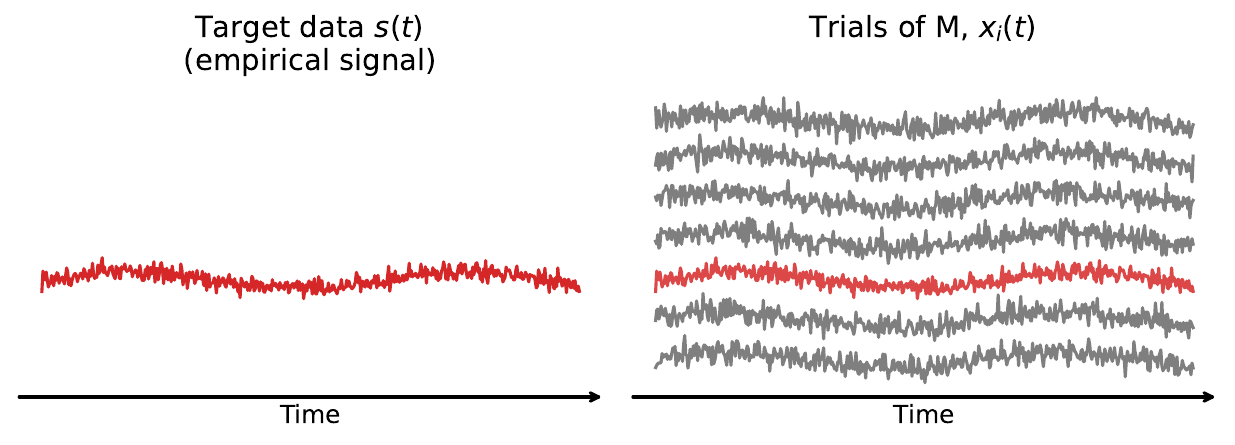}}
 \\
        {\bf B} & {\bf C} & {\bf D}\\
       \includegraphics[scale = 0.5]{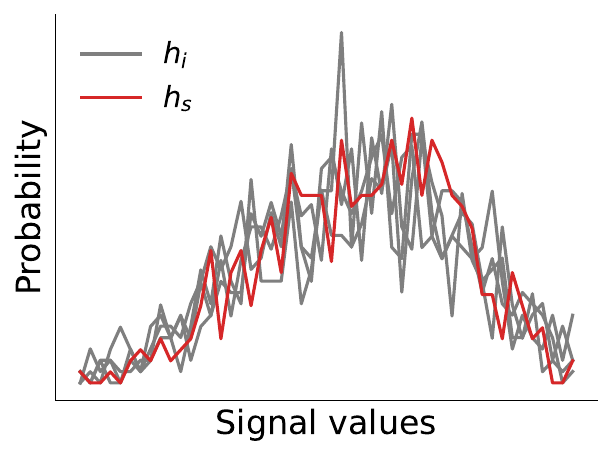} & \includegraphics[scale = 0.5]{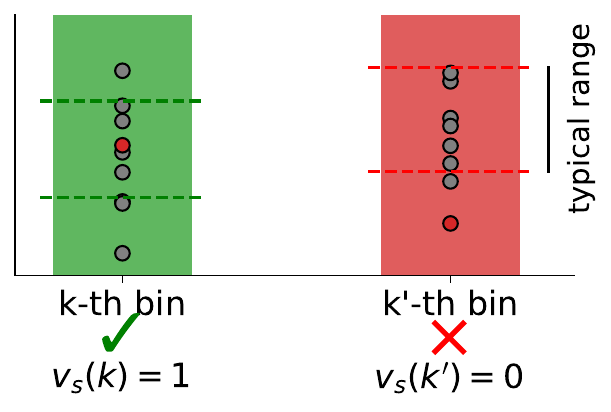} & \includegraphics[scale = 0.5]{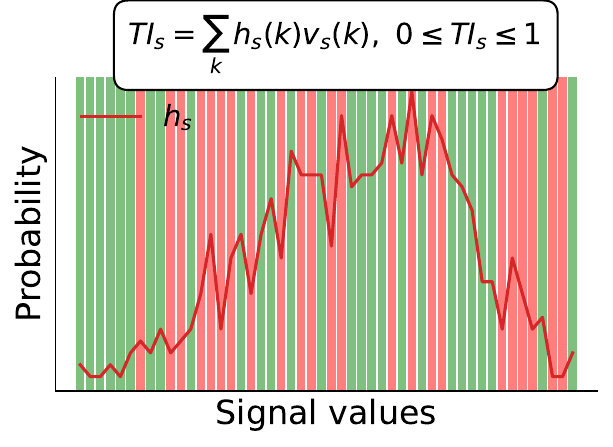}\\
        {\bf E} &  & \\
        \multicolumn{3}{l}{\includegraphics[scale = 0.5]{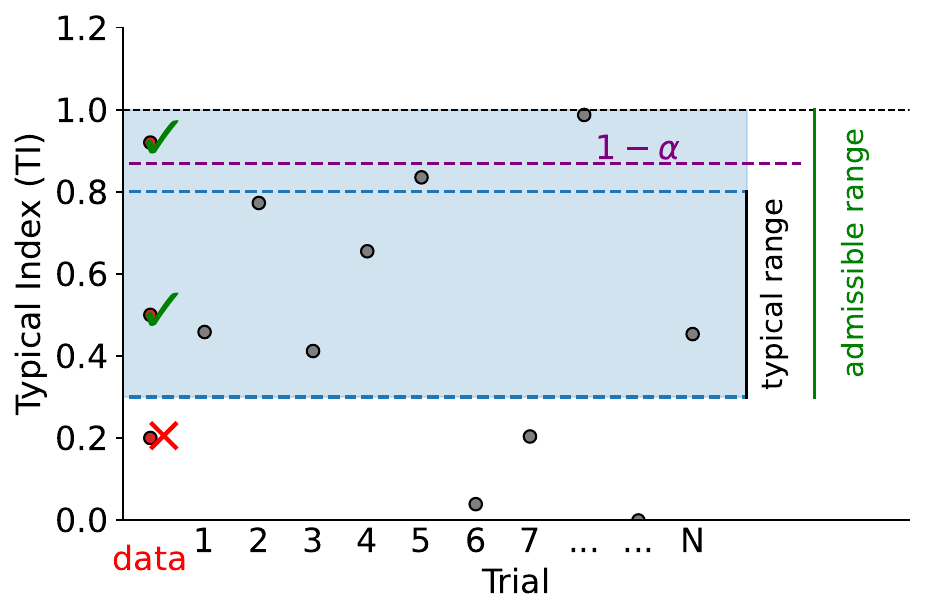}}\\
    \end{tabular}
    \caption{\textbf{Typicality test (stage 1):} Assessing how well  a given (target) time series $s(t)$ blends in (is embedded) within a set of (trial) time series $\{x_i(t)\}_{i=1}^n$ generated by a model \( M \). \textbf{A:} Representative examples of the target (red) and trials (gray) time series.  \textbf{B:} 
 Histograms for the target (\(h_s\)) and trial ($\{h_i\}_{i=1}^n$) time series. The histograms are computed over the same bin range, have the same number of bins (\(K \)) and are normalized by the total area under the histogram curve.
  \textbf{C, D:} For each bin \( k \), we perform Tukey's test to assess whether the value $h_s(k)$ is typical (green) or an outlier (red) from the set $\{h_i(k)\}$.
 We repeat the process for each trial.   \textbf{E:} We compute the typicality index TI and establish whether. If TI is greater than the lower bound from the IQR test or the value $1 - \alpha$ -where \( \alpha \) is a tolerance threshold for deviations-, the typicality test is considered positive.}
    \label{tip_test}
\end{table}

\begin{table}[htpb!]
    \centering
    \begin{tabular}{lll}
        {\bf A} & {\bf B} & {\bf C}\\
       \includegraphics[scale = 0.425]{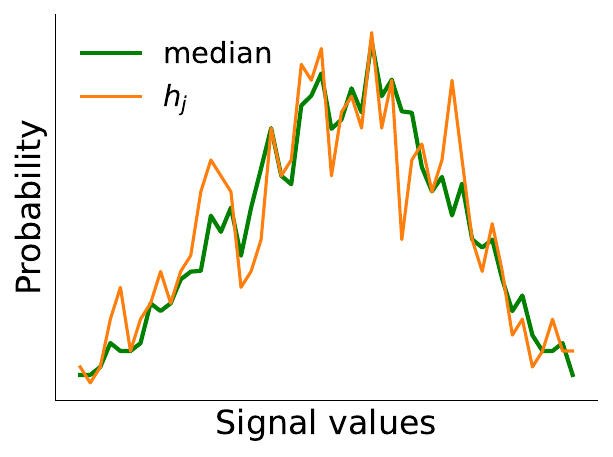} & \includegraphics[scale = 0.425]{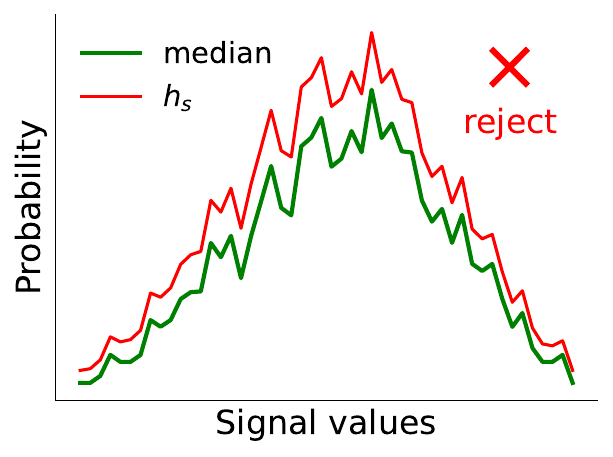} & \includegraphics[scale = 0.425]{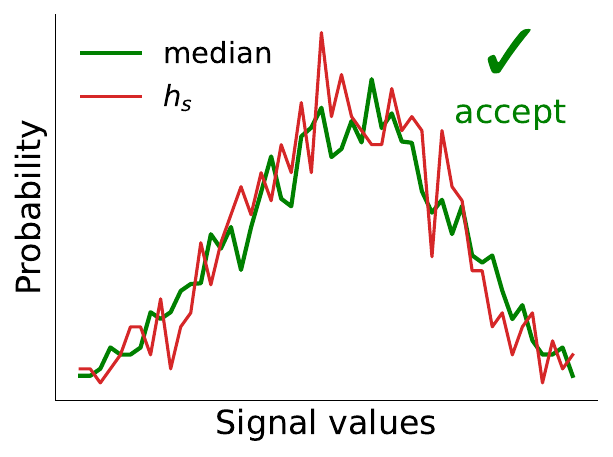}\\
    \end{tabular}
    \caption{\textbf{Sign test for histogram $h_s$ (stage 2):} {\bf A} The histogram for a given trial (generated by the model \( M \); orange) exhibits fluctuations around the median of the set of histograms for all the trials considered (green). If the target time series $s(t)$ has also been generated by the  model \( M \), then the histogram $h_s$ should display similar fluctuations around said median. 
{\bf B.} Example of a histogram \( h_s \) that may pass the typicality test (stage 1), but is rejected by the sign test. 
{\bf C.} Example of a histogram \( h_s \) that passes  the typicality test (stage 1) and is accepted by the sign test.}
    \label{sign_test}
\end{table}

\begin{table}[H]
\begin{center}
\begin{tabular}{c}
\begin{tikzpicture}[
  level distance=2cm,
  level 1/.style={sibling distance=6.5cm},
  level 2/.style={sibling distance=3.5cm},
  box0/.style={rectangle, draw,dashed, rounded corners, line width=1.5pt, align=center, minimum width=3.5cm, minimum height=1cm},
  box1/.style={rectangle, draw, rounded corners, align=center, minimum width=3.5cm, minimum height=1cm},
  box2/.style={rectangle, draw, rounded corners, align=center, minimum width=2.5cm, minimum height=1cm},
  tipico/.style={box1, fill=gray!30, line width=1.5pt, draw=gray!80!black},
  atipico/.style={box1, fill=red!30, line width=1.5pt, draw=red!80!black},
  regular/.style={box2, fill=green!30, line width=1.5pt, draw=green!70!black},
  irregular/.style={box2, fill=orange!40, line width=1.5pt, draw=orange!80!black}
]

\node[box0] {Target (empirical) data $s(t)$}
  child {
    node[atipico] {Atypical}
      child[draw=none] {node[below=1mm, red!80!black] {\shortstack{Inadequate \\ model (\(M\))}}}
  }
  child {
    node[tipico] {Typical}
      child {
        node[irregular] {Irregular}
          child[draw=none] {node[below=1mm, red!80!black] {\shortstack{Inadequate \\ model (\(M\))}}}
      }
      child {
        node[regular] {Regular}
          child[draw=none] {node[below=1mm, green!50!black] {\shortstack{ Adequate \\ model (\(M\))}}}
      }
  };

\end{tikzpicture}
\end{tabular}
\end{center}
\caption{{\bf Adequacy/Reliability (AR) Test: summary.} The target signal is compared against the set of system trials generated by a putative reconstruction model \( M \) 
and categorized as typical/atypical and regular/irregular, resulting in an overall classification of the model \( M \) as  either adequate or inadequate as a reconstruction of the dynamical system \( S \) that generated the target data \( s(t)\).}
\label{test_summary}
\end{table}

\subsubsection{The two stages of the AR Test: Preliminary comments}

The AR test consists of two stages (see Section \ref{twotests}).
The first stage  is an extension of Tukey's interquartile range (IQR) test \cite{tukey1977exploratory}  to histograms in order to neglect outliers by
comparing the geometric structures of \( s(t) \) and each trial in \( x_{trial,n}(t) \) (e.g., Fig.  \ref{tip_test}). 
The second stage uses the sign test to analyze whether there are statistically significant deviations in the observed fluctuations across the surviving trials (e.g., Fig.  \ref{sign_test}). This second analysis is particularly relevant when dealing with noisy signals and stochastic systems. 

We refer to a target trajectory \( s(t) \) that passes  stage 1 as  \emph{typical} with respect to set \( x_{trial,n}\) and we proceed to stage 2. Otherwise this target trajectories are referred to as \emph{atypical} and we conclude that the model \( M \)  is inadequate as a reconstruction of the system \( S \) (Fig. \ref{test_summary}, left). In other words, the target trajectories are not representative of the trajectories produced by the model M.

We refer to a target trajectory  \( s(t) \)  that passes stage 2 (in addition to stage 1) as \emph{regular} with respect to the set \( x_{trial,n}\). Otherwise, we refer to this target trajectory as \emph{irregular} and we conclude that the model \( M \)  is inadequate as a reconstruction of the system \( S \) (Fig. \ref{test_summary}, right).

\subsubsection{Stage 1: typical and atypical target time series}\label{stage1}

This stage evaluates whether the target trajectory \( s(t) \) visits the same regions of the phase-space as the trial trajectories \( x_{trial,n} \) and if it does so  with similar recurrence properties. 

For each bin (\( k = 1, \ldots, K \)), we analyze whether the value \( h_s(k) \) is either typical or an outlier within the set \( h_{trial,n}(k) \)
(see Fig. \ref{tip_test}). We then construct a vector \( v \), where the elements \( v_k \) are defined as \( v_k = 0 \) if \( h_s(k) \) is an outlier or \( v_k = 1 \) if it falls within the range of typical values. In the latter case, we say the bin is typical.
With this information, we define the typicality ($TI_s$-) index   for $h_s$ as the area of this histogram after the outliers have been removed

\begin{equation}
TI_s = \sum_{k=1}^K h_s(k)v_k, 
\end{equation}
which takes values between 0 (all outliers) and 1 (all typical values). 

We repeat this calculation to compute  $TI_i$-index for each trial histogram $h_i \in h_{trial,n}$ with respect the remaining histograms  ($\{h_j,j\neq i\}$), and take the vector $TI_M =:\{TI_i\}_{i=1}^n$. We then assess whether the $TI_s$ index exceeds either the lower bound from the IQR test (computed from the $TI_M$ set) or the value $1 - \alpha$, where \( \alpha \) is a  threshold value (determined by the user), using the less stringent of the two as a reference. If it does,  we proceed with the next stage of the test. Otherwise, we  conclude \( h_s \) is not representative of the set of trial histograms and abort the AR test at this stage. 

\subsubsection{Stage 2: Regular and irregular target time series }\label{stage2}

This stage assesses whether the local fluctuations observed in \( h_s \) are similar to those observed in \( h_{trial,n}\) and therefore reflect the same dynamics as the trials generated by the model \( M \).

We evaluate the degree to which the histogram \( h_s(k) \) deviates from the median curve \( m(k) \) for each \underline{typical} bin (with \( v_k = 1 \)). We chose the median because it  represents the central tendency of the set \( h_{trial,n} \).  More specifically, we analyze whether these deviations are consistent with what would be expected under a binomial distribution. To this end, we apply a sign test  \cite{conover1999practical} with a significance level \(\alpha\) (in order to test the null hypothesis \( H_0 \) that the observed deviations are compatible with a binomial distribution). For each typical bin (\(k\)), we compute the median \( m(k) \) of the set \(\{h_i(k)\}_{i=1}^n\). We then conduct the binomial test to verify  \( H_0\)
and compare the  \( \text{p-value} \) with \( \alpha \) (see Fig. \ref{sign_test})

\begin{itemize}
\item If \( \text{p-value} > \alpha \), we do not reject \( H_0 \) (with significance $\alpha$) and conclude that the reconstruction of the geometric structure is adequate with a typicality index \( TI \).  
\item If \( \text{p-value} < \alpha \), we reject \( H_0 \) and conclude that the reconstruction is inadequate.
\end{itemize}

\noindent Note that having a good approximation of $m(k)$ requires a sufficiently large number of trials \( x_i(t) \).

\subsubsection{Interpretation of the parameter \(\alpha\)}
The parameter \( \alpha \) is used as a threshold in both stages of the AR test. In the second stage, \( \alpha\) represents the probability of rejecting the null hypothesis when it is in fact true that the histogram \(  h_s \) of the target time series \( s(t) \) follows the distribution represented by the system’s median histogram. More generally, \(\alpha\) should be understood as a measure of the strictness with which statistical compatibility is required between the target trajectory and the collective properties observed in the set of model trials.
Note that the value of \( \alpha \)  does not represent the probability of a false negative (i.e., rejecting the hypothesis that the trajectory was generated by the system when that hypothesis is actually true). 

\subsection{Implementation of the AR test}

\subsubsection{Validation: Stochastic Systems with Underlying Chaotic Dynamics}

The reconstruction chaotic dynamics in the absence of stochastic components (e.g., additive noise) has been extensively addressed in recent literature~\cite{koppe2019identifying,durstewitz23,ozalp2023reconstruction}. 
Sensitivity to initial conditions makes direct comparison between chaotic time series and their putative reconstruction of limited utility. This has motivated the use of alternative approaches, such as histogram comparison (see Section \ref{geometricstructure}), to ensure that the long-term behavior of the two signals is consistent. However, as discussed in Section \ref{geometricstructure}, when the system is stochastic,
direct comparison between the geometric structures of the system's trajectory and its reconstruction, which use histograms, become an inadequate strategy.

The  Lorenz model (see Section \ref{models}) is a prototypical example of a chaotic system that has often been used as a case study to evaluate the effectiveness of  proposed methods~\cite{hemmer2024optimal}. 
Figure~\ref{fig_lorenz}-A illustrates a realization of the stochastic Lorenz model (simulated data; blue) and a reconstructed trajectory (model trial; orange) having the same initial condition. 
The reconstruction was obtained by using modified PLRNN (see Section \ref{models}). The added hidden layer consisted of 80 nodes and noise manually introduced in the network.
As expected, due to the influence of chaotic deterministic component and stochastic nature of the underlying system, the two trajectories do not coincide except possibly for small number of points (Figure~\ref{fig_lorenz}-A1 to A3). However, they tend to visit the same regions of phase space (Figure~\ref{fig_lorenz}-A4).

To validate the reconstruction using the two-stage AR test, we generated 120 trajectories from the trained PLRNN (i.e.,the putative reconstruction), using the same initial conditions as the original (target) data. We then constructed the histograms for the entire set of time series corresponding to the variable $x$. The histograms covered the range of values of all time series and the number of bins was determined according to the rule $\#\textrm{bins} = \sqrt{\textrm{lenght of data}}$. The same procedure was applied to the variables $y$ and $z$.

The Kullback-Leibler divergence between the histograms of the simulated and reconstructed data across trials shows a wide dispersion for all three variables  (Figure~\ref{fig_lorenz}-B). As discussed in Section \ref{geometricstructure}, this lack of robustness renders the metric inconclusive to assess the adequacy of the reconstruction. This remains true for the other metrics considered in Section \ref{geometricstructure} (Figure~\ref{fig_lorenz_aux} in Appendix~\ref{inadequacy_appendix}).

In contrast, the AR test (with a factor $f = 1.5$ for the IQR stage  and significance level $\alpha = 0.05$) produced a positive result, indicating that the simulated (target) data is both typical and regular compared to the 100 trials simulated using the reconstructed system and therefore the reconstructed system is adequate (in all variables, the t-index of the data was greater than $0.95$). 

Figure~\ref{fig_lorenz}-C shows, for each variable $x$, $y$, $z$, a comparison between the histograms of the target data, selected trials from the trained PLRNN system, and the median curve (computed across trials), which is used in the second stage of the AR test.

We can observe -mainly in the histograms of the \(x\) variable, and to a lesser extent in those of the \(y\) variable- that the target data histogram does not adhere as closely to the median as the net trials do. However the fluctuations of data target histogram around this median are sufficient to successfully pass the second step of the test. For the \(z\) variable, in contrast, a good agreement is observed between the target data histogram and the median curve of the set of trials.

\begin{table}[H]
\begin{tabular}{ll}
{\bf A1} & {\bf A2}\\ 
\includegraphics[width=0.325\textwidth]{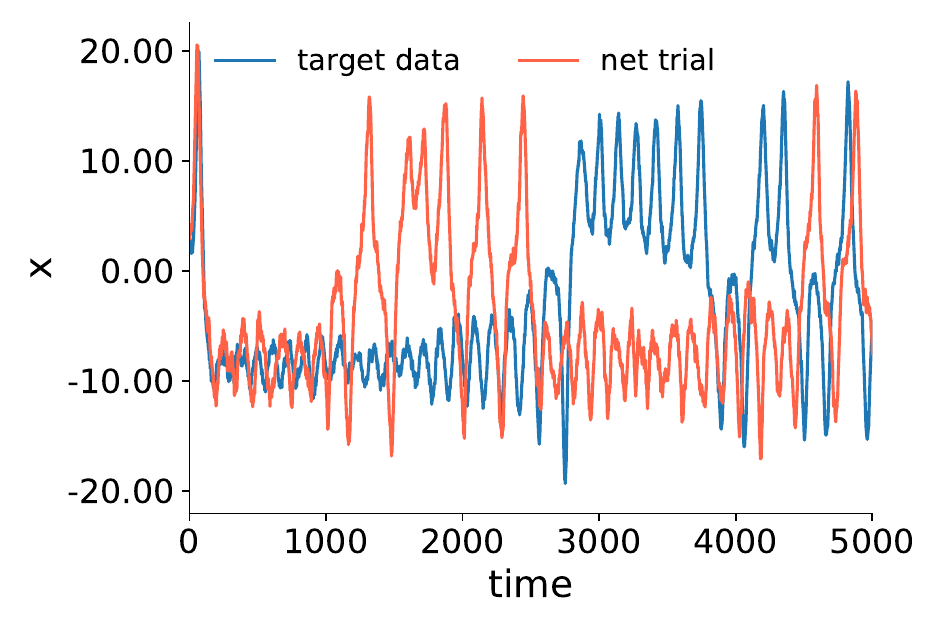} &
\includegraphics[width=0.325\textwidth]{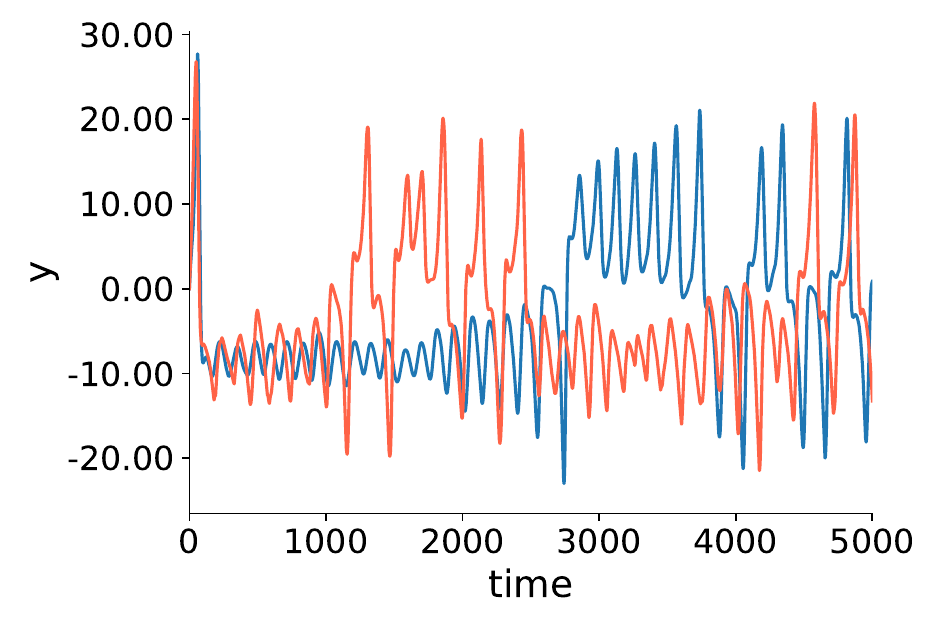}\\
 {\bf A3} & {\bf A4}\\
\includegraphics[width=0.325\textwidth]{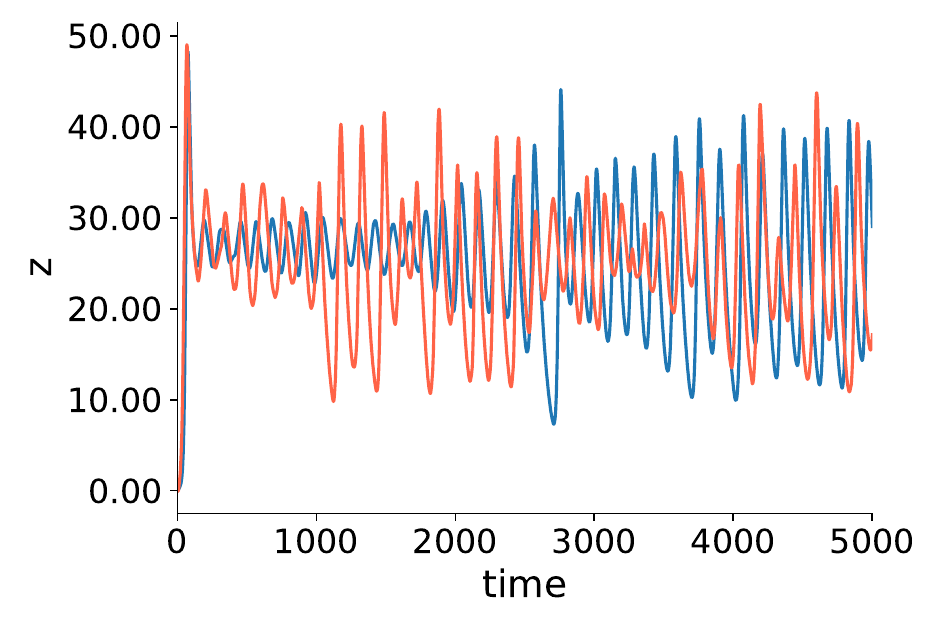} & \includegraphics[width=0.325\textwidth]{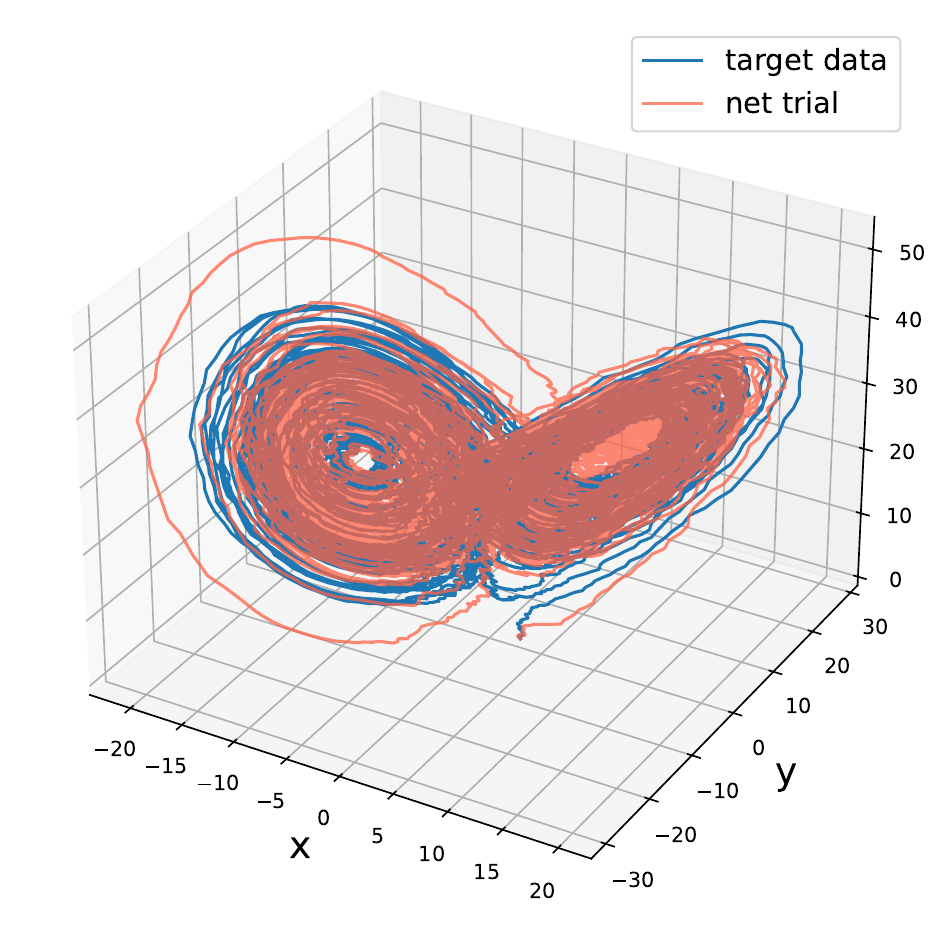} \\
\end{tabular}\\
\begin{tabular}{lll}
{\bf B1} & {\bf B2} & {\bf B3} \\
\includegraphics[width=0.325\textwidth]{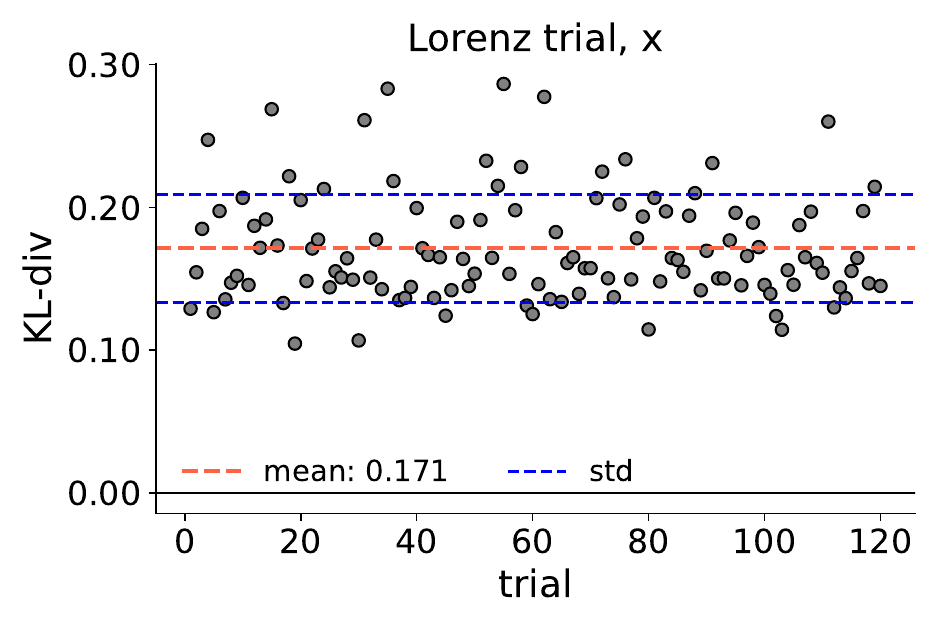} &
\includegraphics[width=0.325\textwidth]{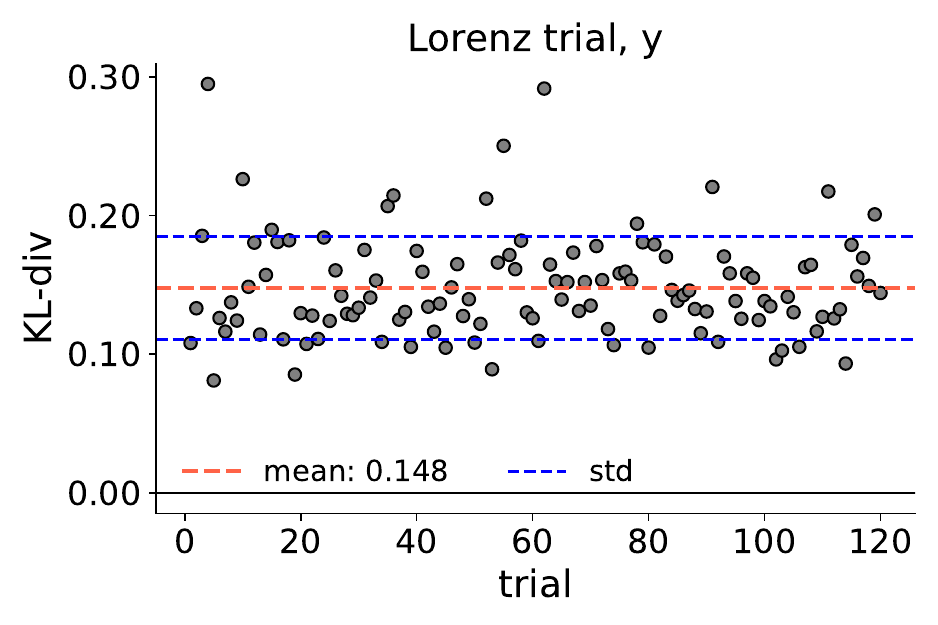}&
\includegraphics[width=0.325\textwidth]{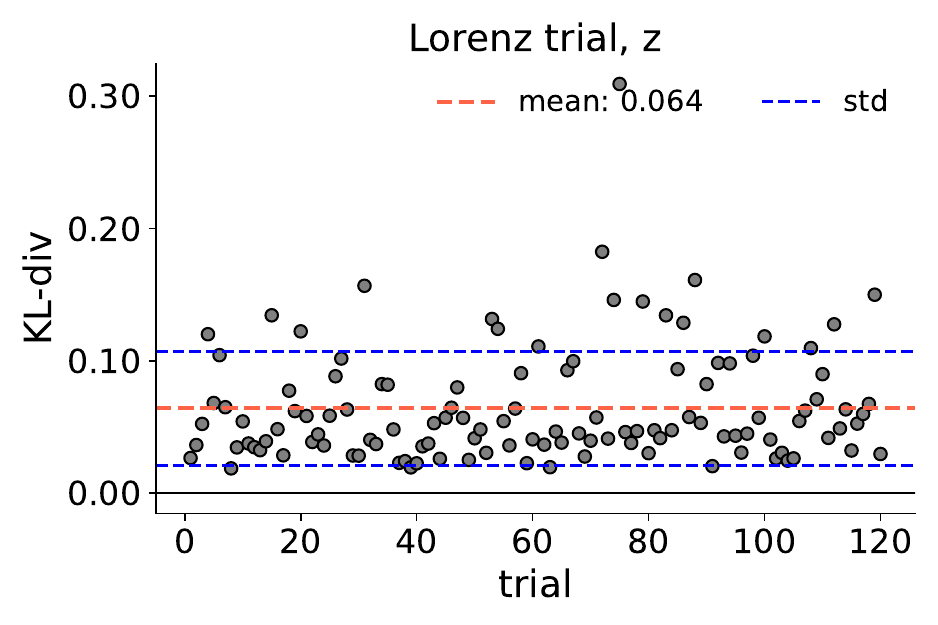} \\
{\bf C1} & {\bf C2} & {\bf C3}\\
\includegraphics[width=0.325\textwidth]{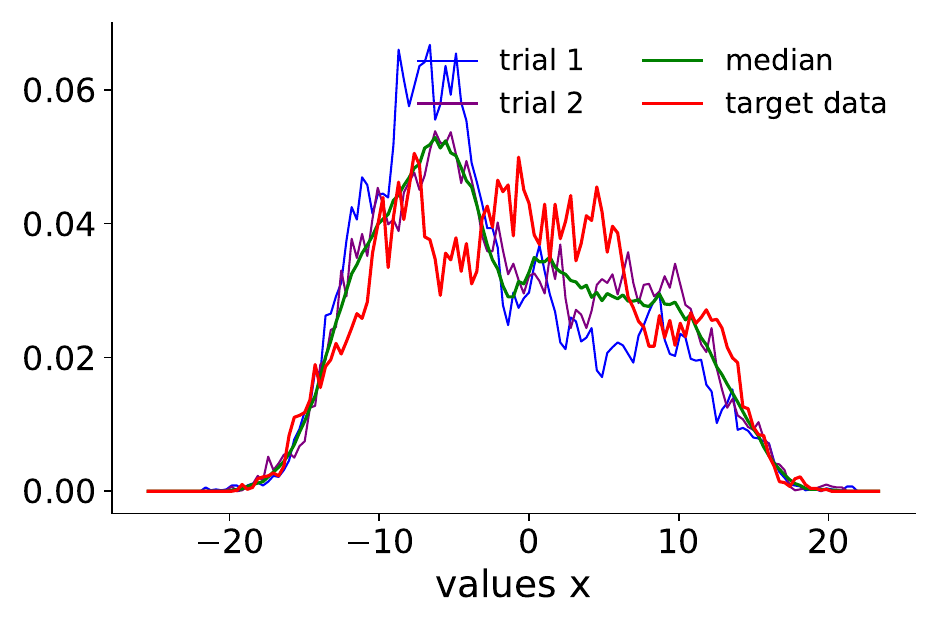} &
\includegraphics[width=0.325\textwidth]{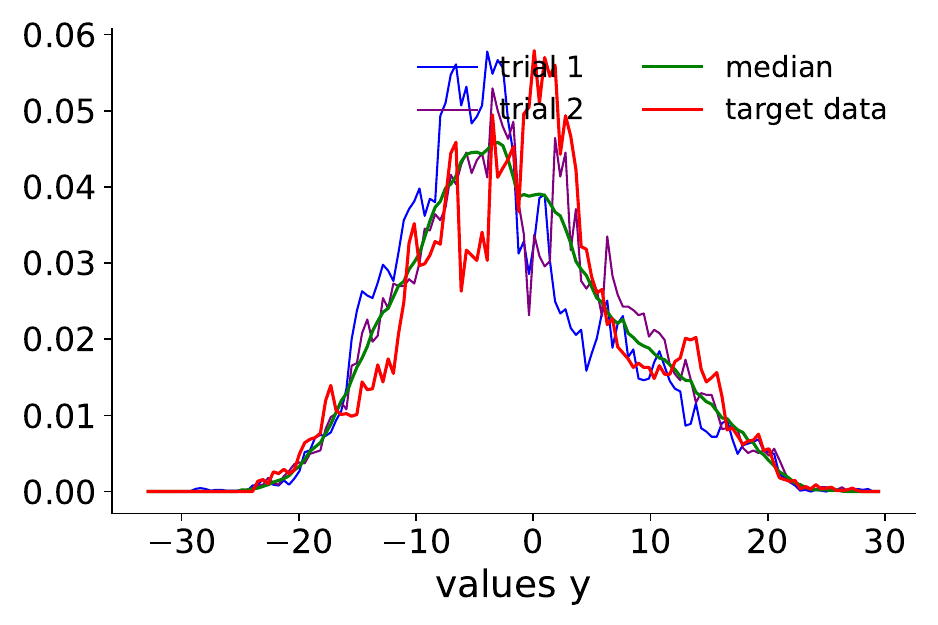}&
\includegraphics[width=0.325\textwidth]{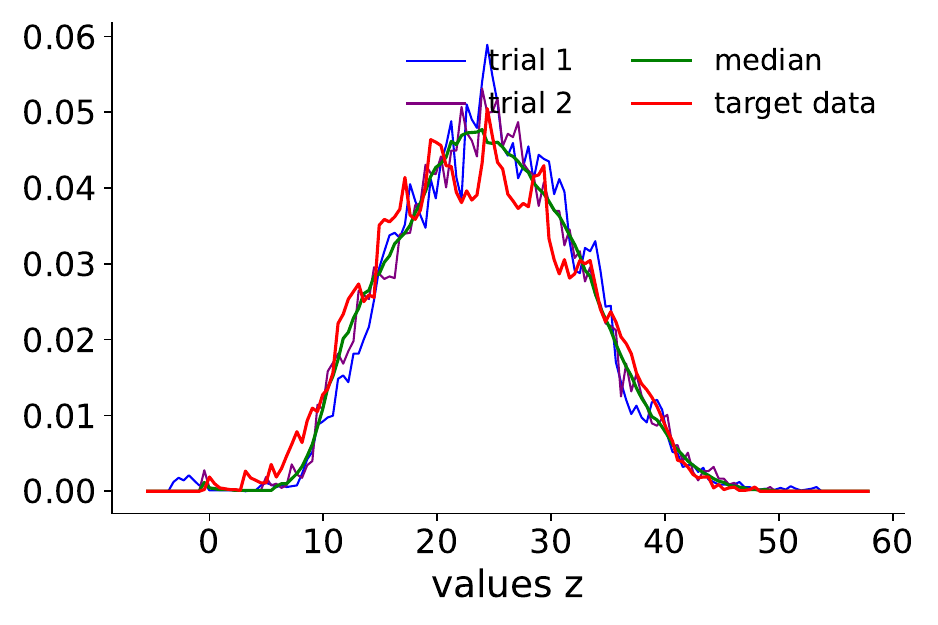} \\
\end{tabular}
\caption{\small {\bf Reconstruction of the stochastic Lorenz system.} 
{\bf A.} Representative example of the target data for the simulated Lorenz system (red) and a single reconstruction trial (blue) using a PLRNN. {\bf A1- A3} Time series. {\bf A4} Phase-space diagram illustrating the strange attractor. 
{\bf B.} Kullback–Leibler divergence between the histograms for the target data and trials as a function fo the trial number for the three model variables. 
{\bf C.} Superimposed graphs of the median over all trials (green), histograms of the target data (red) and histograms of two arbitrary trials (blue and purple) for the three model variables. We used the following parameter values: $\sigma = 10$, $\rho = 28$, $\beta = 8/3$ and $D= 15$. The system was simulated using a time step of $\Delta t = 0.005$ and a total of $N = 20000$ integration steps.
}
\label{fig_lorenz}
\end{table}

\subsubsection{Stability of test/convergence of the exploratory procedure/consistency across increasing numbers of trials}

When analyzing the adequacy of a putative reconstruction/model for a given target data, the comparison is done between these target data $s(t)$ and a set of simulated trials from the system $x_{n,trial}(t)$, which is expected to be representative of its dynamics within a large enough region of interest.  The larger the set of simulated trial data, the more reliable test is expected to be since more information is available about the system's dynamics. However, the appropriate size of this set to be ``representative enough"  depends on the system itself and cannot be determined \textit{a priori}.

We analyze whether the outcome of the proposed exploratory AR test converges as the size of the set of trials increases. We refer to this property as the "stability of the test outcome", or simply as the "stability of the test".

In Figure~\ref{test_sta}, we analyze the stability of the AR test outcome as the size of the set of trials $x_{n,trial}(t)$ is varied. To this end, one trial of the system (fixed as the target data) is selected, and the AR test is applied using sets of trials of increasing size. The analysis is repeated on the same system under different noise levels. In all cases, the test is performed using only the time series corresponding to the first variable ($x$ variable).

Figures~\ref{test_sta}.A and~\ref{test_sta}.B show the results of the stability analysis for the Lorenz and FitzHugh-Nagumo systems, respectively (the first has a chaotic deterministic part and the second has a periodic orbit as its only stable attractor). For the Lorenz system, it can be observed that, at low and middle noise levels ($D=5$, $10$, $30$), the test correctly identifies the system, and the outcome is not affected by the size of the trial set. As the noise level increases ($D=50$), different outcomes appear as the trial set grows; however, these group into well-defined ``bands,'' indicating that the test outcome does not change intermittently. Furthermore, for the high noise level ($D=50$), the outcome is less stable, which can be attributed to the difficulty of identifying the system under intense noise conditions (see also Figure~\ref{test_sta_orbit}.A in Appendix~\ref{test_sta_noise} for an illustration of the effect of different noise levels).

On the other hand, for the FitzHugh-Nagumo system, the test exhibits greater stability across all noise levels considered. Under moderate noise ($D = 0.05$), a negative outcome occurs but later it stabilizes around a positive result. For the highest noise level, the results become less stable and show perturbations for longer trial sets. However, once again, this can be attributed to the difficulty of finding comparable trials when the system is subjected to strong fluctuations due to noise (see Figure~\ref{test_sta_orbit}.B in Appendix~\ref{test_sta_noise} for an illustration of the effect of different noise levels). Even so, in this case, although false negatives (``inadequate'') do occur, they always arise from the second step of the test, while the experimental trial is consistently evaluated as ``typical''.

\begin{table}[H]
\begin{tabular}{ll}
{\bf A} & \includegraphics[width=0.85\textwidth]{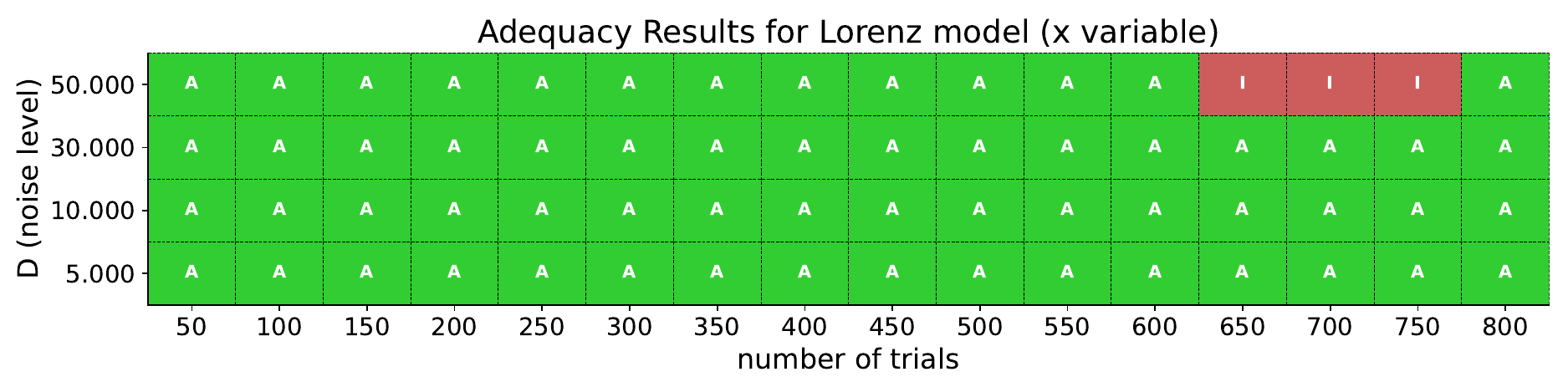}\\
{\bf B} & \includegraphics[width=0.85\textwidth]{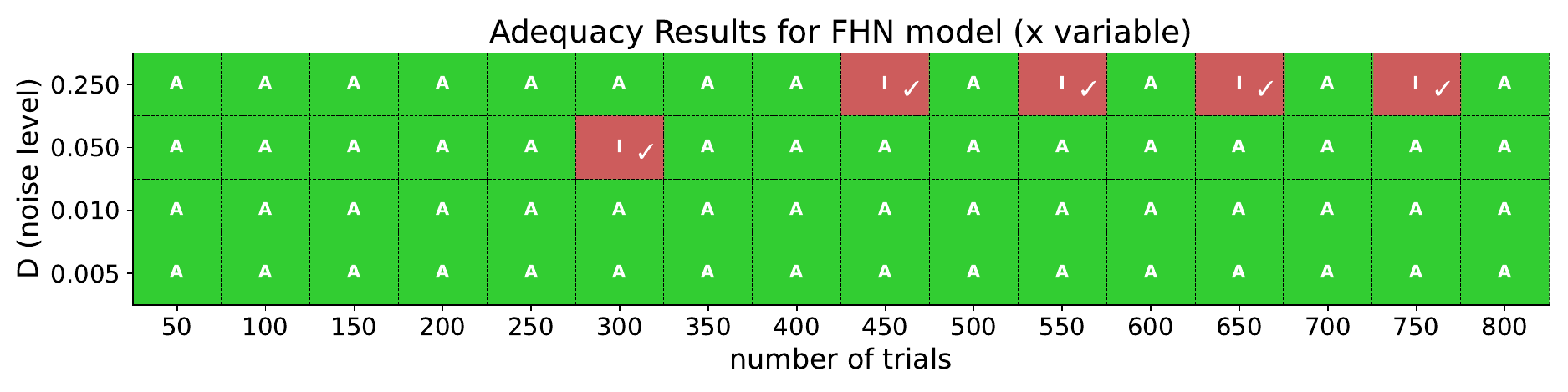} \\
\end{tabular}
\caption{{\bf Analysis of stability of AR test outcome.} {\bf A.} Results of applying the AR test to a trial of the Lorenz system using the system itself, as the size of the trial set $x_{n,trial}(t)$ increases. Each trial was run for 20{,}000 steps, $\Delta t = 0.005$. The system parameters are the same as those in Figures~\ref{variability_problem} and \ref{fig_lorenz}. {\bf B.} The corresponding results for a FitzHugh-Nagumo system. Each trial was run for 10{,}000 steps, $\Delta t = 0.025$. The system parameters are the same as those in Figure~\ref{variability_problem}. The color of each cell indicates the test result: green for adequate (``A") and red for inadequate (``I"). In the latter case, the tilde denotes whether the outcome of the first step of the test was 'typic' (tilde) or 'atypic'.}
\label{test_sta}
\end{table}

\subsection{System reconstruction meets system degeneracy: effects on AR test outcomes}

\emph{Identifiability} refers to the ability of uniquely determine the model parameters from the measured or observed data in a process that evolves according to the laws proposed by that model. Consequently, even if the system is subject to stochastic fluctuations, the resulting data allow for a relatively reliable estimation of its parameters. \emph{Unidentifiability}, the lack of identifiability,  is associated with the notion of \emph{degeneracy} where different combinations of parameter values can produce the same observable output. Thus, the degeneracy of a system is not related to the statistical uncertainty arising from data collection, but it is embedded in the mathematical structure of the models.  Identifiability and unidentifiability, along with their implications for biological systems, have been extensively studied across a wide range of models \cite{bellman1970structural, cobelli1980parameter, audoly2002global, raue2009structural,lederman2022parameter} (see in particular \cite{lederman2022parameter} for a comprehensive review of the literature).

To analyze how the degeneracy of a system (and the consequent unidentifiability) affects the implementation of the test, we resort to two typical systems used to study and illustrate the different mechanisms that can give rise to unidentifiability: the so-called $\Lambda\Omega$ model and the FitzHugh-Nagumo (FHN) model.

\paragraph{$\Lambda\Omega$ model:}{
The model is given by equations \eqref{lo_eq}. For \( \lambda > 0 \), the stable limit cycle is a circle centered at the origin. 
A comprehensive study of this system is presented in \cite{lederman2022parameter}. The $\Lambda\Omega$ model admits several parameter combinations that share the same (stable) limit cycle  with radius $\bar{r} =\sqrt{\lambda / b}$, traversed with the same frequency $\bar{\omega} = \omega+a\, \bar{r}$, and differ only in the transient dynamics of decay toward the attractor (see Figure \ref{LO_system}.B), which critically depends on \( \lambda \).

Therefore, the ability to discriminate between degenerate $\Lambda\Omega$ systems based on their orbits will depend on how different the degenerate parameters are and  on the level of noise that activates the transient dynamics.

We consider a reference $\Lambda\Omega$ system with fixed parameters \((a=\omega=1 \text{ and } \b=\lambda=0.1)\), subjected to different levels of additive noise in the first variable (\(x\)), controlled by the parameter \(D\) (see Section \ref{noise_protocol}). For each value of \(D\), a trial is generated and used as the target data corresponding to that noise level. Then, for the same value of \(D\), we vary the parameters \(\lambda=b\), while keeping both the condition \(\lambda / b  = 1\) and the remaining parameters fixed in order to preserve both the stable limit cycle and its frequency. Finally, we apply the AR test between the target series associated with that value of \(D\) and a set of 500 trials generated by the system with modified values of \(\lambda\) and \(b\). In this way, we evaluate whether the AR test can correctly identify the system that originated the target data, as well as discriminate systems that do not coincide with the reference one.

Figure \ref{LO_system}.A shows the stochastic orbits used as target data where it can be observed that excessive levels of noise ($D=0.5$) disrupt the expression of oscillatory and regular behavior. Panel \ref{LO_system}.C presents the results of the AR test: for moderate/high noise levels (the three intermediate values of $D$), the system was always correctly identified when compared with itself. However, lower noise level ($D= 0.001$) prevented self identification: the AR test produces a false negative, although the negative test result occurred at the second step (the typicality test was passed).

It is also observed that, for moderate noise levels ($D = 0.025$, $0.005$), the AR test classifies systems that are very similar to the original as "Adequate", leading to false positives, showing that the systems are not yet sufficiently different to be distinguished, or that the transients have not been sufficiently activated to differentiate them. These false positive results disappear either when the values of the parameters $\lambda$ and $\beta$ are increased and the systems become more different from each other, or when the noise level is increased.

Moreover, for very low noise levels, all results are negative while the typicality test is still passed, except when the systems are sufficiently different (i.e., when their parameters differ enough). Rather than reflecting efficient identification, this seems to indicate the impossibility of obtaining stable patterns in the resulting orbits under very low noise levels, which would allow for effective comparison.

Furthermore, in Appendix \ref{LO_others}, additional results from the same procedure are presented, showing that in the presence of systems with strong structural degeneracy, the test does not produce stable outcomes.
}

\begin{table}[H]
\begin{tabular}{c c}
\multicolumn{2}{c}{
\begin{minipage}[t]{0.98\textwidth}
\textbf{A}\\[2mm]
\includegraphics[width=\linewidth]{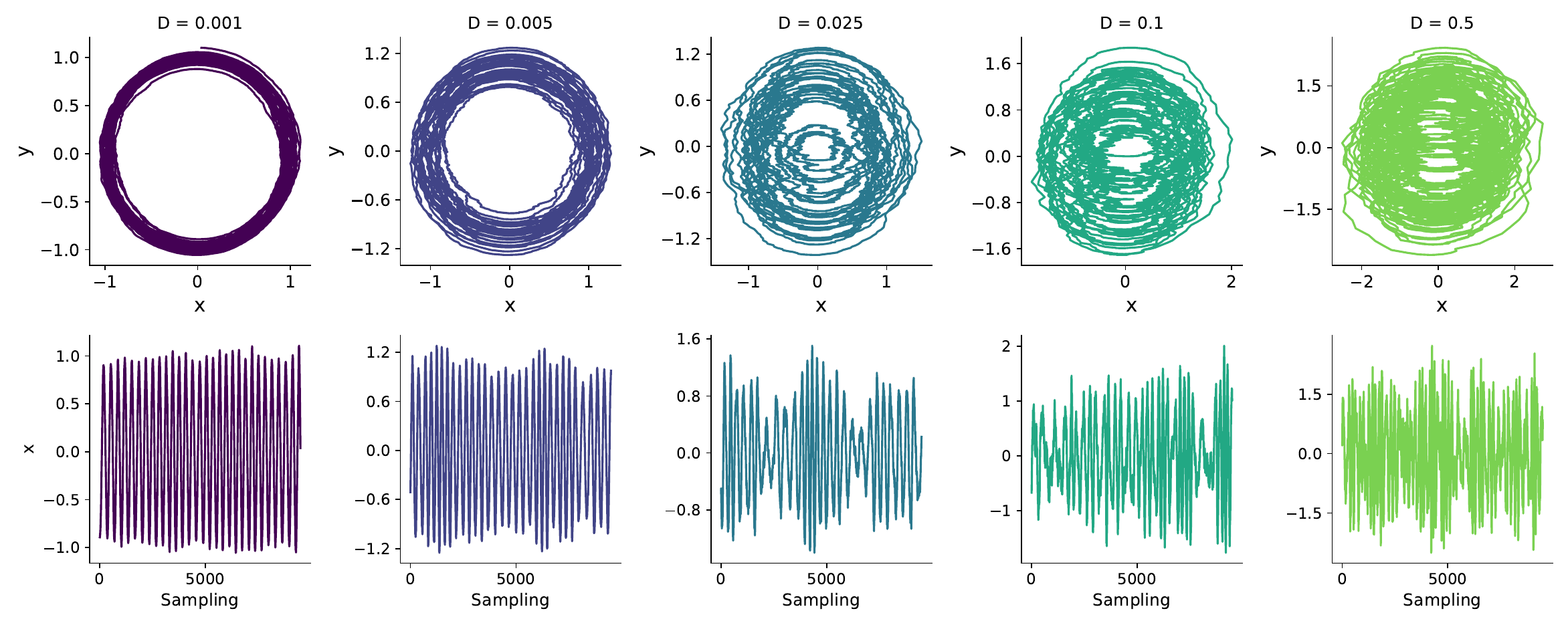}
\end{minipage}
}\\[2mm]
\begin{minipage}[t]{0.35\textwidth}
\textbf{B}\\[2mm]
\includegraphics[width=\linewidth]{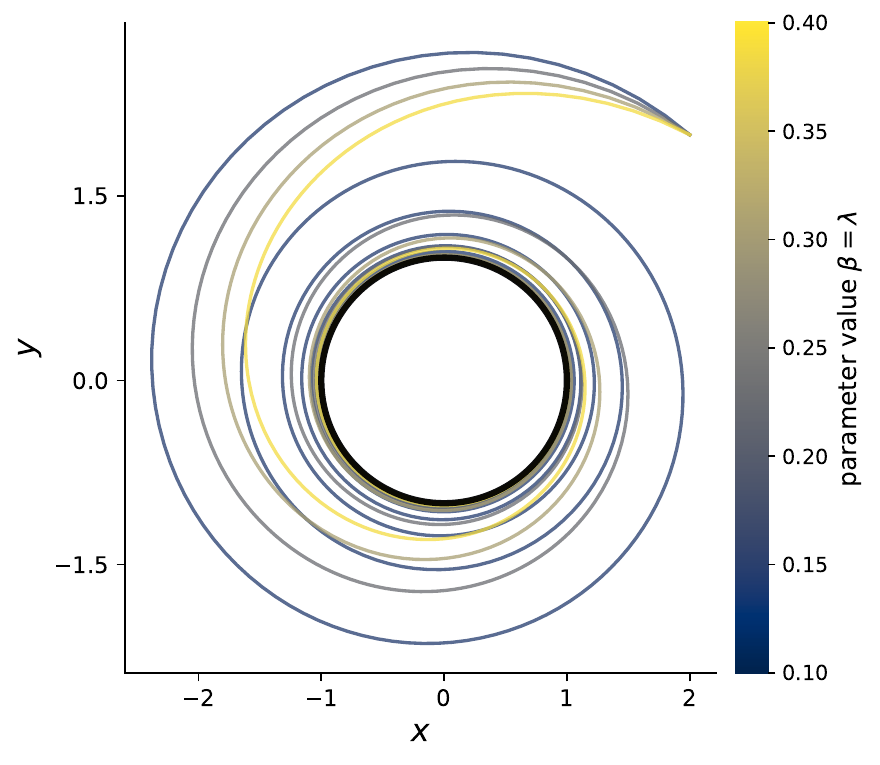}
\end{minipage}
&
\begin{minipage}[t]{0.65\textwidth}
\textbf{C}\\[2mm]
\includegraphics[width=\linewidth]{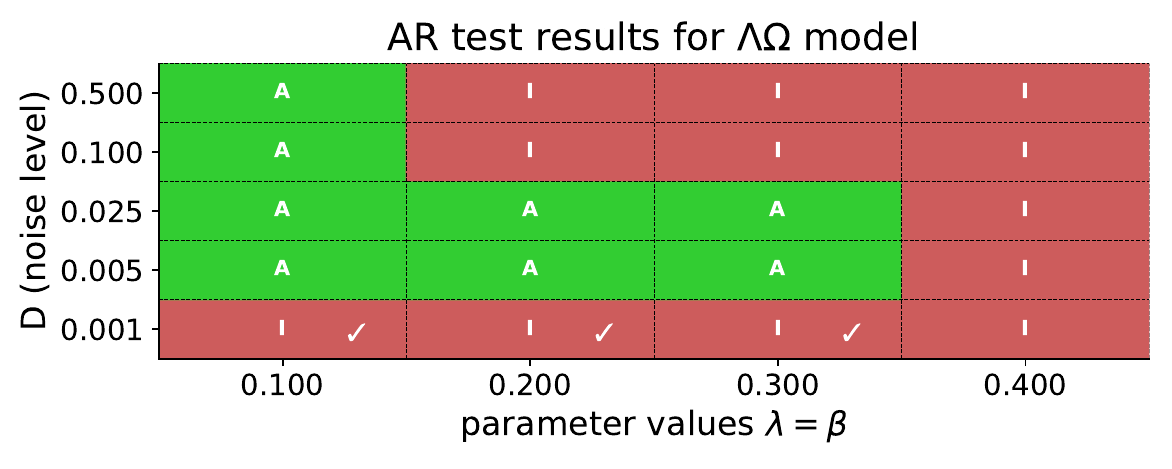}
\end{minipage}\\
\end{tabular}
\caption{
\textbf{Analysis of the identification of the $\Lambda\Omega$ model using the AR test.} {\bf A.} Target series generated for different levels of additive noise (see \ref{noise_protocol}) applied to $x$ variable, the reference system is defined by the following parameter values $\lambda = \beta = 0{.}1$, $\omega = \alpha = 1$. These orbits are used as target series for the AR test. {\bf B.} Transient dynamics of the model when keeping $\alpha$ and $\omega$ fixed, and varying $\beta$ and $\lambda$ (while always maintaining $\beta = \lambda$). All considered systems share the same limit cycle with identical frequency. {\bf C.} Results of applying, for each noise level $D$, the AR test to the target series corresponding to the reference system and to a set of 500 trials generated by the system with parameter values $\lambda = \beta$.
Green cells correspond to results classified as ``adequate''(A) and red cells to ``inadequate''(I); the check mark inside the red cells indicates that the result was typical, meaning the system was classified as inadequate in the second step of the test. For the trial simulations, a time step of $\Delta t = 0.01$ was used, and $N = 10{,}000$ integration steps were performed.}
\label{LO_system}
\end{table}

\paragraph{FitzHugh-Nagumo (FHN) model:}{The model is given by \eqref{fhn_eq}. The FHN model, in turn, exhibits different parameter combinations that produce oscillations with similar amplitudes or frequencies. In particular, we consider variations in the parameter $a$ (with other parameters fixed) that give rise to a subcritical Hopf bifurcation \cite{strogatz2024nonlinear,izhikevich2007dynamical}. Consequently, in the presence of noise and near the bifurcation values, the phenomenon of stochastic resonance may occur, leading to oscillations with a frequency shared with that of the oscillatory system \cite{gammaitoni1998stochastic,dykman1998can,moss1993stochastic,pei1995detection}.

We performed the same analysis as for the $\Lambda\Omega$ system, but varying only the parameter $a$ while keeping the others fixed. We begin with $a=-0.5$, where there is a stable focus close to the bifurcation. In this way, the considered systems are not topologically equivalent, although, due to the phenomenon of stochastic resonance in the presence adequate level of noise, the orbits retain some similarity, see Figures \ref{FHN_system}.A and \ref{FHN_system}.B.
 
Although the FHN system exhibits parameter sets in which certain observable attributes (amplitude, frequency, or duty cycle) remain very similar, it does not show structural degeneracy as pronounced as the $\Lambda\Omega$ model, so the AR test distinguishes different systems more effectively. In Figure \ref{FHN_system}.C, the first column shows a correct identification of the system for all noise levels; however, this result should be qualified. For low or moderate noise levels ($D = 0.005 - 0.025$), the correct identification is maintained even when the parameters are varied slightly. In contrast, in this same scenario, for high noise levels, the AR test produces false positives. Nevertheless, when the systems differ significantly from each other, the identification is correct in all scenarios.
}

\begin{table}
\begin{tabular}{c c}
\multicolumn{2}{c}{
\begin{minipage}[t]{0.98\textwidth}
\textbf{A}\\[2mm]
\includegraphics[width=\linewidth]{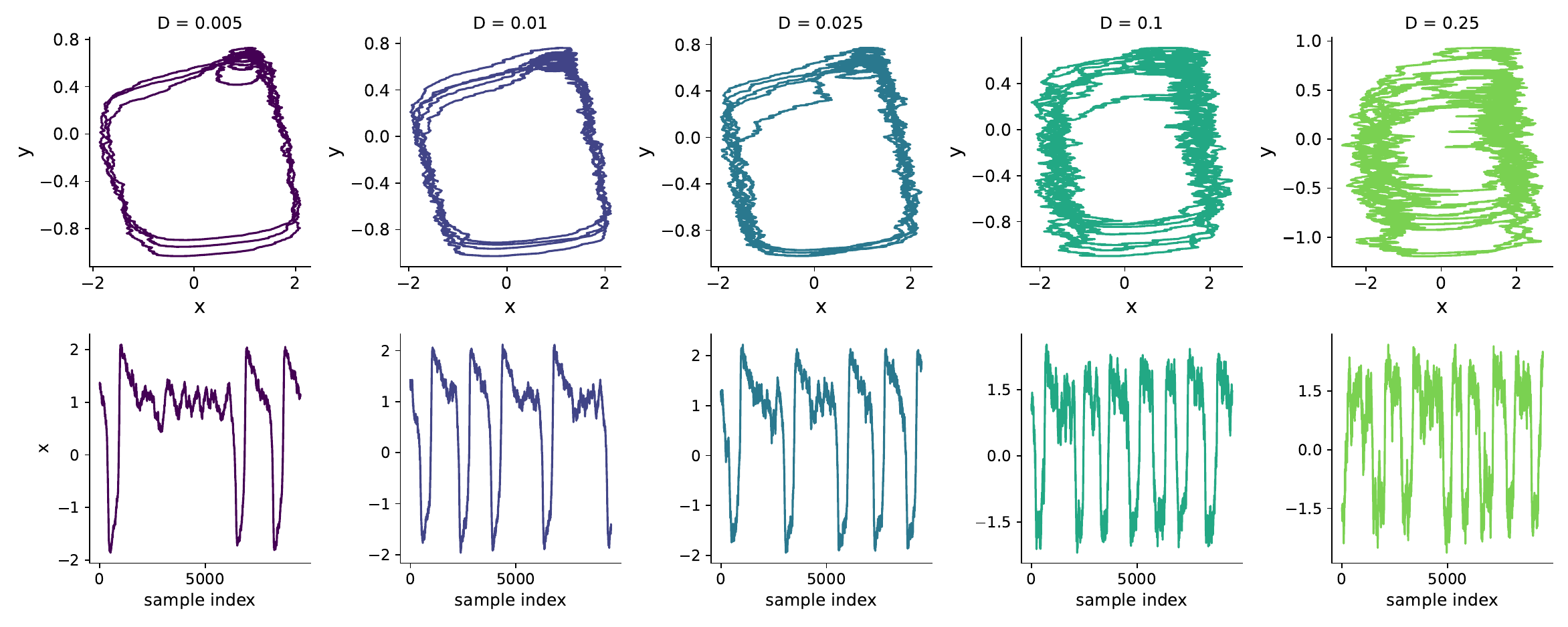}
\end{minipage}
}\\[2mm]
\begin{minipage}[t]{0.35\textwidth}
\textbf{B}\\[2mm]
\includegraphics[width=\linewidth]{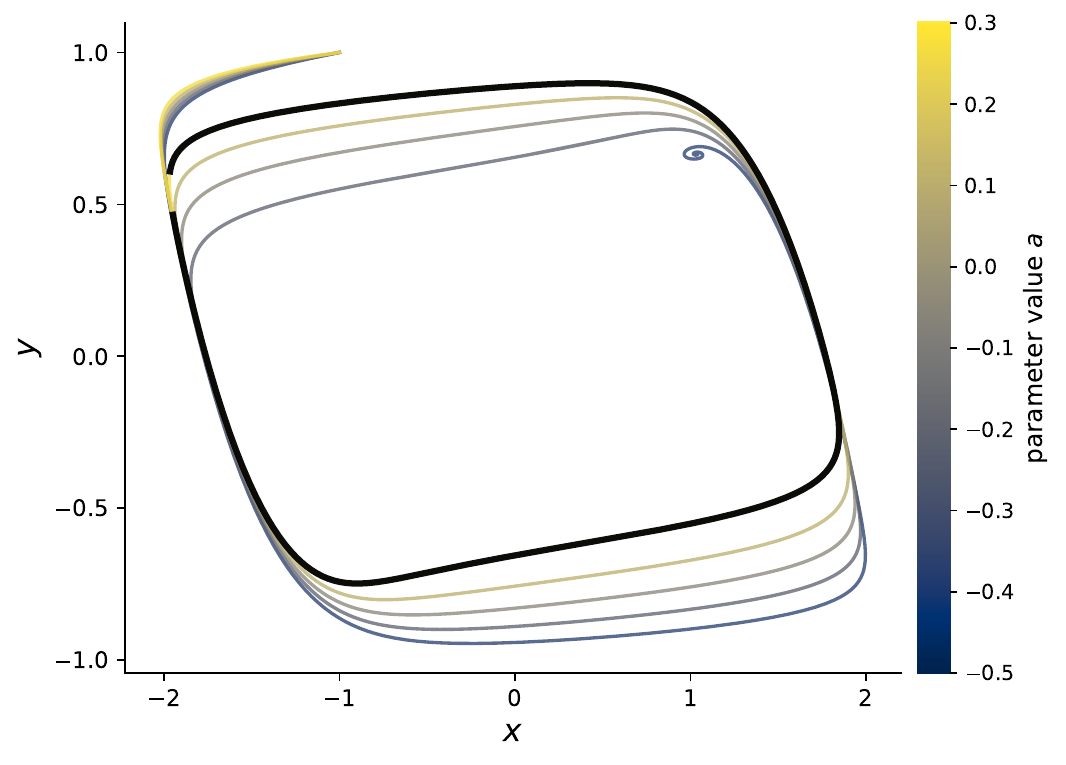}
\end{minipage}
&
\begin{minipage}[t]{0.65\textwidth}
\textbf{C}\\[2mm]
\includegraphics[width=\linewidth]{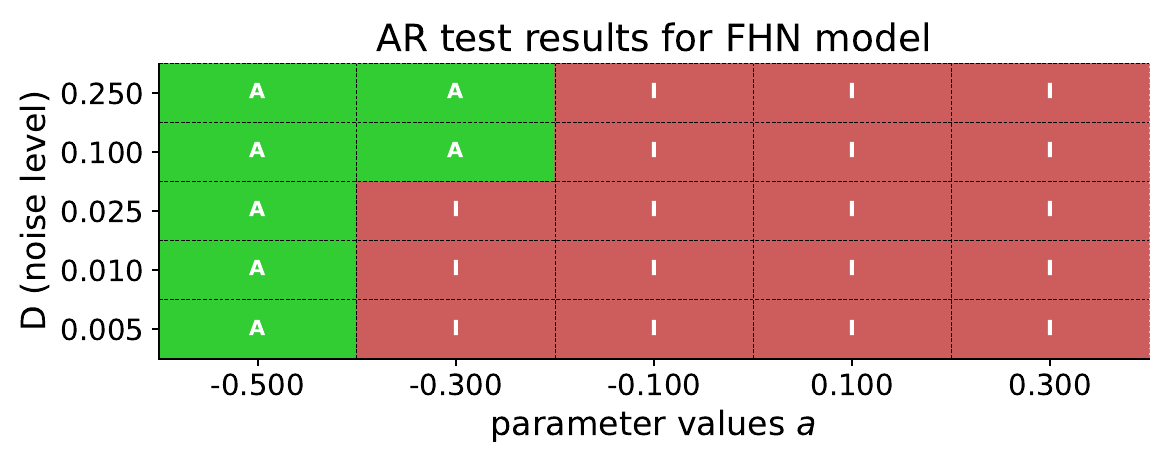}
\end{minipage}
\end{tabular}
\caption{\textbf{Analysis of the identification of the FHN model using the AR test.} The same analysis performed for the $\Lambda\Omega$ model was applied here, but in this case varying the parameter $a$, which controls the excitability of the system. Other parameter values: $b = 0.8$, $\tau = 12$, $I=0$. {\bf A.} Effect of noise in the FHN system taken as reference ($a=-0.5$). {\bf B.} Different orbits obtained by varying the parameter $a$. {\bf C.} Results of apply the AR test for each noise level $D$. For the trial simulations, a time step of $\Delta t = 0.025$ was used, and $N = 10{,}000$ integration steps were performed.}
\label{FHN_system}
\end{table}

\subsubsection{Effects of reconstruction-induced biases}

\begin{table}
\begin{tabular}{c c}
\multicolumn{2}{c}{
\begin{minipage}[t]{0.98\textwidth}
\textbf{A}\\[2mm]
\includegraphics[width=\linewidth]{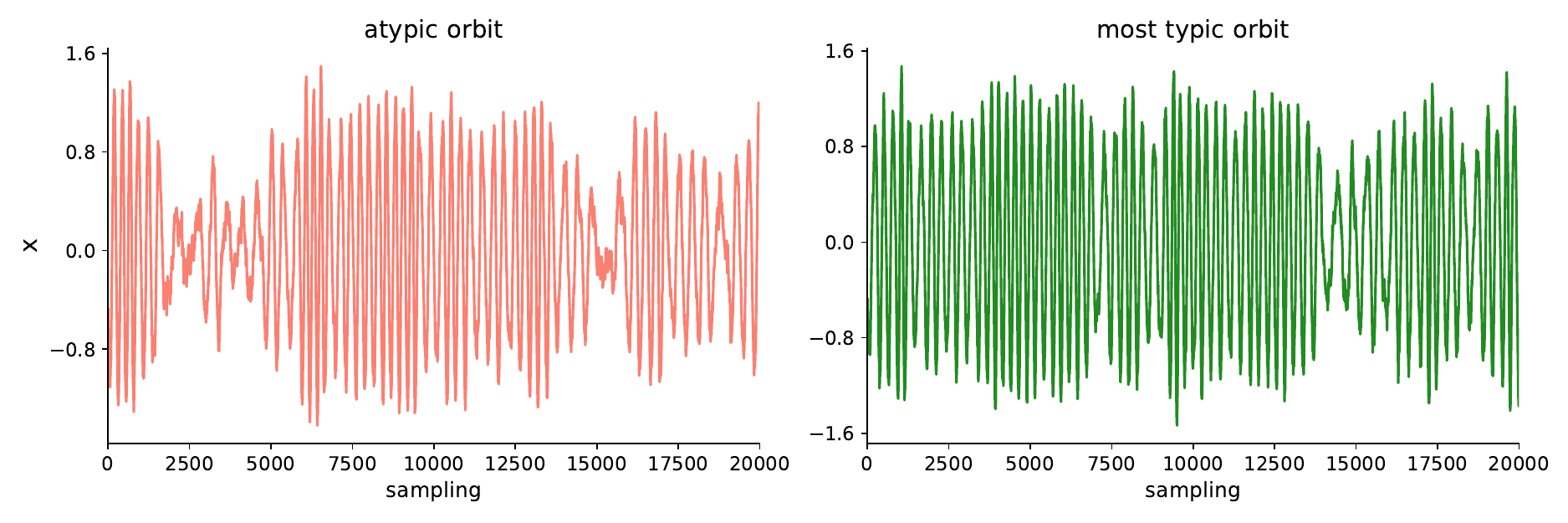}
\end{minipage}
}\\[2mm]
\begin{minipage}[t]{0.45\textwidth}
\textbf{B}\\[2mm]
\includegraphics[width=\linewidth]{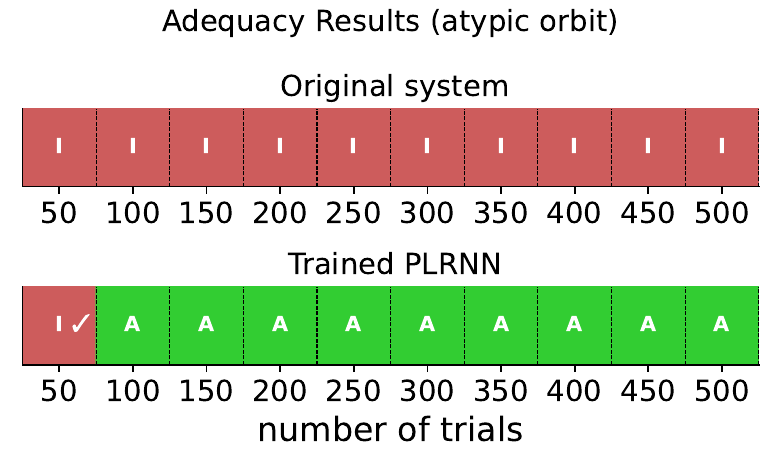}
\end{minipage}
&
\begin{minipage}[t]{0.45\textwidth}
\textbf{C}\\[2mm]
\includegraphics[width=\linewidth]{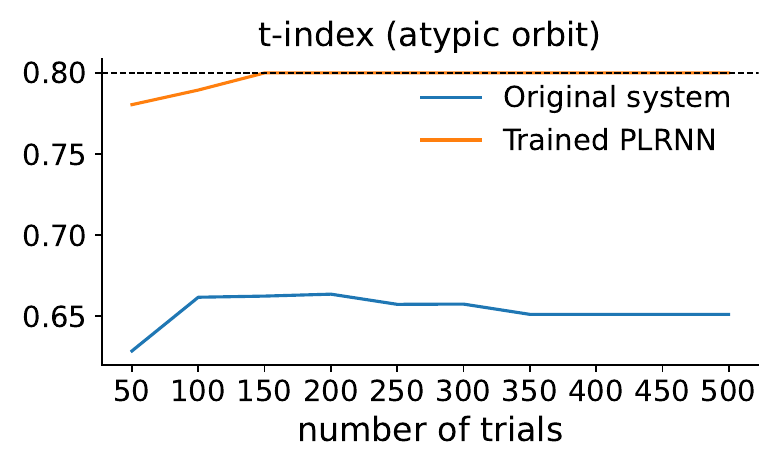}
\end{minipage}
\end{tabular}
\caption{\textbf{Preference of the test for the putative reconstruction over the original system.} 
{\bf A.} The atypic orbit (red) selected as target data from a set of 300 orbits generated by the $\Lambda\Omega$ model and the one with the highest t-index (the \emph{most typical orbit}, in green). {\bf B.} Result of the adequacy test for different sizes of the comparison trial set. {\bf C.}Value of the \emph{t}-index of the atypical orbit for different trial-set sizes. The consistently high values, close to 0.8, indicate the network’s ability to reproduce the patterns present in the training target data. Parameter values: $\lambda = b = 0.1$, $a = 1$, $\omega = 1$, $D = 0.2$; for trial simulations we used a time step $\Delta t = 0.01$ and $20{,}000$ steps.}
\label{sysnet}
\end{table}

Each trial of a stochastic system exhibits two types of patterns: on the one hand, those governed by the system’s own dynamics (though perturbed by noise), and on the other, \emph{spurious} patterns, which correspond to fluctuations arising solely from noise variability. At the same time, the loss of uniqueness characteristic of stochastic systems implies that a trial may be typical (``standard / common / representative") or, alternatively,  ``exceptional'', which is possible, but rare  (for instance, a higher proportion of spurious patterns compared to those intrinsic to the system). In systems with a high degree of degeneracy (such as the $\Lambda\Omega$ systems presented above), this may lead to reconstructions based on poorly representative orbits being classified as ``adequate'' by the AR test, while the same test applied to the original system yields an ``inadequate'' result.

This is particularly relevant because, in general, reconstruction methods do not discriminate between intrinsic and spurious patterns present in the target data, but instead use all available data to generate a system that replicates the observed dynamics. Consequently, a trial produced by a putative reconstruction may reproduce, in a more balanced proportion, the various patterns present in the target data than the original system itself. This gives rise to a new type of ``artificial" (
``induced'' / ``trained'') degeneracy: starting from an unusual trial of a system, it is possible (for instance, through the training of a network) to generate another system for which the selected training orbit becomes suitable, yet more representative than in the original system.

To illustrate the above, we again consider the $\Lambda\Omega$ model. We generated a set of 300 orbits using a reference $\Lambda\Omega$ system ($\omega=\alpha =1$, $\lambda = \beta = 0.1$) with additive noise in the first variable ($D=0.2$). We then evaluated the \emph{t}-index of each orbit with respect to the set and selected as target data an atypical orbit, namely, an orbit whose \emph{t}-index lies outside the typical range (see \ref{stage1} and Figure \ref{sysnet}.A). Using this orbit, we trained a PLRNN with 80 hidden nodes, to which noise was manually added afterward.

In panel \ref{sysnet}.B, we present the results of applying the AR test using the selected atypical orbit as \emph{target data} and considering trial sets of different sizes. In all cases, the AR test yields a negative result (false negative) for the $\Lambda\Omega$ system that generated the target data, whereas the result is predominantly positive (false positive) for the trained PLRNN. Moreover, in all cases the \emph{target data} is considered atypical when compared with the trials generated by the $\Lambda\Omega$ system, while it is always classified as ``typical'' when compared with the trials generated by the network. In Fig.~\ref{sysnet}.C, we show the evolution of the t-index of the target data, which is systematically larger when computed with respect to sets of trials generated by the network than when computed with respect to trials generated by the original system.

\section{Discussion}

 We have address the problem of reconstructing stochastic dynamics from noisy data. Rather than focusing on the reconstruction algorithms, we tackle the less-studied problem of properly validating these reconstructions. The common approach of minimizing a loss function during training and using that as a criterion for ``sufficient approximation'' has the drawback of depending too strongly on the chosen loss function and on the specific dataset being reconstructed. We showed that the acceptable error thresholds vary significantly depending on the type of data and on additional assumptions about the systems generating the dynamics (whether they are chaotic, exhibit stable equilibria, etc.). Moreover, we showed that the intrinsic variability of stochastic systems also makes standard methods for analyzing the quality of reconstructed deterministic dynamics inaccurate in this context, in addition to reintroducing the problem of defining acceptable error thresholds, which are highly case-dependent.

To overcome these issues, we proposed a test (the two-stage AR test) to assess the \emph{adequacy} of a reconstruction using a \emph{blending} approach: to determine whether a reconstruction is adequate, we propose analyzing how recurrent or representative the noisy experimental trial is within a set of trials generated by the reconstructed system. This test has an exploratory  character (in the sense of exploratory data analysis). It provides a flexible, model-independent validation tool that can be naturally integrated into the reconstruction process, complementary to the tools for the reconstruction itself.
Unlike other validation approaches that require splitting the available data into training and validation sets, our method allows for the use of the \emph{entire dataset} during the reconstruction stage. The test can then be applied \emph{a posteriori}, without compromising the amount of information used for model fitting. This makes it particularly suitable in scenarios where data are limited, as it enables a \emph{more optimal use of the available dataset}.

If two competing models result to be adequate according to this test, then an additional metric will be required to select one over the other. 
Unlike the approaches discussed earlier, this method is exploratory and universal and can also be used to compare reconstructions obtained using different methods and based on different datasets.

\bigskip
\noindent
{\bf Scope and Interpretation of the Test}
\medskip

We emphasize that the AR test is \emph{not meant to classify the reconstruction provided by the model M as ``good'' or ``bad'' in an absolute sense}, but rather in relation to the statistical and dynamical structure of the reference set~$x_{trial,n}$. In fact, as discussed earlier, the AR test provides a way to assess whether a given target  is well embedded into (``blends in") the global geometry of the set. In other words,  whether the target trajectory aligns with or deviates from the ``manifold" formed by the ensemble of reference trajectories.

This perspective is particularly useful when comparing reconstructions generated using different methods or based on distinct datasets, as it provides a \emph{unified criterion} to assess the degree of consistency or deviation from a reference behavior, without relying on method-specific assumptions.

\bigskip
\noindent
 {\bf Conceptual generality of the approach}
\medskip

We also emphasize that behind the AR test there is a general two-step approach, which must be distinguished from the specific form or tools we used to implement it:
\begin{enumerate}
    \item The first step of our approach analyzes whether the target data is adequately ``aligned'' around the same ``manifold'' (point cloud) as the trials generated by the putative system.
    
    \item The second step, given that the match is never perfect, analyzes whether the fluctuations of such ``alignment'' are consistent with those observed within the set of trials of the putative system.
\end{enumerate}

In this work, we implemented the first step by generalizing the IQR test to a set of histograms, whereas the second step was implemented through the analysis, via a sign test, of the differences between the histogram of the target data and the median curve of the histograms from the set of trials. However, our general approach remains open to alternative implementations of these two steps: another formalization of the IQR test, or a different analysis of the recurrence relationships in the observed phase space, could lead to a different implementation of the first step, while alternative estimators to the median or other statistical tests could be employed to analyze the fluctuations in the second step. Future research should address these aspects of the general AR test.

\bigskip
\noindent
{\bf Exploratory Nature of the AR Test and its minimal assumptions}
\medskip

While the proposed test has statistical components, collectively,  it is an exploratory test (as opposed to a statistical test) since it does not require the prior specification of context-dependent tolerance thresholds. 
The only two parameters involved are the factor $f$ used in the IQR test and the parameter $\alpha$, both of which are set to standard values commonly found in the literature: $f = 1.5$ and $\alpha = 0.05$. Rather than relying on a case-specific calibration, the test provides a general and flexible framework for identifying deviations of the target trajectory from he set of trial trajectories (generated by the putative model M) that may warrant further analysis. This makes it suitable for application across a wide range of contexts. Specifically, the test evaluates whether the variability of the experimental data remains within certain admissibility ranges, which are constructed individually for each case according to a general rule. The width of these ranges depends on the specific characteristics of the system $M$ that generates the sampling of trials.

It is worth noting that this test —being a generalization of Tukey's test for the case of histograms— does not require setting error tolerance thresholds, which are usually defined according to the application context and the system \( M \) under consideration. The determination of the admissibility range in the first step of the test is automatic and depends solely on the variability of the histograms in the set of trials of \( M \). Thus, if the model \( M \) generates a set of trials with greater or lesser variability, the admissibility range (see Fig.\ref{tip_test} E) will automatically adjust, becoming wider or narrower accordingly.

Finally, another noteworthy feature of the test is that it does not require an ergodicity assumption nor any hypothesis about the cardinality of the system \( S \), which can be either continuous or discrete.

\bigskip
\noindent
{\bf Some remarks on the context of degenerate systems}
\medskip

Degeneracy  imposes limits on the application of this test when attempting to distinguish between competing reconstructions, and its results may be affected by the non-identifiability of the underlying dynamics. In the case of continuous systems, degeneracy arises from the possibility that two systems governed by different evolution laws may share a common solution (as in the case of $\Lambda\Omega$ systems), or that their solutions---although distinct---share certain measurable attributes (frequency, amplitude, duty cycle, etc.). In stochastic systems, however, the problem becomes more delicate: the non-uniqueness of solutions, their inherent ``non-repeatability,'' and the ``representativeness'' of selected attributes prevent a straightforward definition of degeneracy (or the guarantee of system distinguishability). As we have shown, systems that are discernible in the deterministic regime may become indistinguishable once noise is added, since the stochastic realizations can exhibit overlapping attributes. In this context, we show that the power of the test to identify the system that generated a given trial depends on the noise level and on how ``similar'' the system may be to others. Moreover, as we demonstrate, the appearance or selection of possible but rare realizations of the system -i.e., trajectories that are not highly recurrent or representative- can bias both the reconstruction of the dynamics and the outcome of the test, leading to situations in which a reconstruction is deemed more ``adequate'' for reproducing certain datasets than the original system itself. Although such scenarios are not the most common, it is important to make them explicit, as they illustrate how degeneracy imposes a theoretical limit on both reconstruction algorithms and their validation approaches.

Finally, it is worth mentioning that the attempt to distinguish systems by using the estimated distributions of their noisy trajectories naturally leads to considering the statistical significance of some classical test (examining the null hypothesis that the systems are identical or not), as well as the verification (or assumption) of hypotheses regarding the ergodicity of the systems, in order to establish the representativeness of the numerically obtained distributions. Our approach shows that such formal statistical analysis is not the only possible route: an exploratory analysis alone already allows one to discriminate between systems, depending on the diversity and the level of activation of the transient dynamics involved. Further research is required to address these issues.

By providing a mechanistic representation of the observed system, the reconstructed model serves as a powerful tool for further analysis, simulation, perturbation, and modification, allowing for deeper insights into its underlying mechanisms.

\paragraph{Acknowledgments}{GC and UC acknowledge support from the Universidad Nacional del Sur grant PGI 24/L131 and CONICET,
Argentina. HGR acknowledges support from the National Science Foundation grants IOS-2002863. HGR is a member of the Mathematical and Computational Neurosciences Collective (MCBC) at NJIT, a Graduate Faculty Member in the Graduate Program in Neuroscience (GPN) in the Center for Molecular and Behavioral Neuroscience (CMBN) at Rutgers University, and a 
Corresponding Researcher at CONICET, Argentina.}

\paragraph{Author Contributions}{ G.C., U.C., and H.G.R. contributed equally to the conception, analysis, and writing of the manuscript.}

\bibliography{biblio}

@article{durstewitz2023reconstructing,
  title={Reconstructing computational system dynamics from neural data with recurrent neural networks},
  author={Durstewitz, Daniel and Koppe, Georgia and Thurm, Max Ingo},
  journal={Nature Reviews Neuroscience},
  volume={24},
  number={11},
  pages={693--710},
  year={2023},
  publisher={Nature Publishing Group UK London}
}

@article{koppe2019identifying,
  title={Identifying nonlinear dynamical systems via generative recurrent neural networks with applications to f{MRI}},
  author={Koppe, Georgia and Toutounji, Hazem and Kirsch, Peter and Lis, Stefanie and Durstewitz, Daniel},
  journal={PLoS computational biology},
  volume={15},
  number={8},
  pages={e1007263},
  year={2019},
  publisher={Public Library of Science San Francisco, CA USA}
}

@article{durstewitz2017state,
  title={A state space approach for piecewise-linear recurrent neural networks for identifying computational dynamics from neural measurements},
  author={Durstewitz, Daniel},
  journal={PLoS computational biology},
  volume={13},
  number={6},
  pages={e1005542},
  year={2017},
  publisher={Public Library of Science San Francisco, CA USA}
}

@inproceedings{monfared2020transformation,
  title={Transformation of ReLU-based recurrent neural networks from discrete-time to continuous-time},
  author={Monfared, Zahra and Durstewitz, Daniel},
  booktitle={International Conference on Machine Learning},
  pages={6999--7009},
  year={2020},
  organization={PMLR}
}

@article{song2016training,
  title={Training excitatory-inhibitory recurrent neural networks for cognitive tasks: a simple and flexible framework},
  author={Song, H Francis and Yang, Guangyu R and Wang, Xiao-Jing},
  journal={PLoS computational biology},
  volume={12},
  number={2},
  pages={e1004792},
  year={2016},
  publisher={Public Library of Science San Francisco, CA USA}
}

@inproceedings{takens2006detecting,
  title={Detecting strange attractors in turbulence},
  author={Takens, Floris},
  booktitle={Dynamical Systems and Turbulence, Warwick 1980: proceedings of a symposium held at the University of Warwick 1979/80},
  pages={366--381},
  year={2006},
  organization={Springer}
}

@article{cybenko1989approximation,
  title={Approximation by superpositions of a sigmoidal function},
  author={Cybenko, George},
  journal={Mathematics of control, signals and systems},
  volume={2},
  number={4},
  pages={303--314},
  year={1989},
  publisher={Springer}
}

@inproceedings{brenner2023multimodal,
  title={Multimodal teacher forcing for reconstructing nonlinear dynamical systems},
  author={Brenner, Manuel and Koppe, Georgia and Durstewitz, Daniel},
  booktitle={When Machine Learning meets Dynamical Systems: Theory and Applications},
  year={2023}
}

@article{kramer2021reconstructing,
  title={Reconstructing nonlinear dynamical systems from multi-modal time series},
  author={Kramer, Daniel and Bommer, Philine Lou and Tombolini, Carlo and Koppe, Georgia and Durstewitz, Daniel},
  journal={arXiv preprint arXiv:2111.02922},
  year={2021}
}

@article{hemmer2024optimal,
  title={Optimal Recurrent Network Topologies for Dynamical Systems Reconstruction},
  author={Hemmer, Christoph J{\"u}rgen and Brenner, Manuel and Hess, Florian and Durstewitz, Daniel},
  journal={arXiv preprint arXiv:2406.04934},
  year={2024}
}

@article{lederman2022parameter,
  title={Parameter estimation in the age of degeneracy and unidentifiability},
  author={Lederman, Dylan and Patel, Raghav and Itani, Omar and Rotstein, Horacio G},
  journal={Mathematics},
  volume={10},
  number={2},
  pages={170},
  year={2022},
  publisher={MDPI}
}

@article{bellman1970structural,
  title={On structural identifiability},
  author={Bellman, Ror and {\AA}str{\"o}m, Karl Johan},
  journal={Mathematical biosciences},
  volume={7},
  number={3-4},
  pages={329--339},
  year={1970},
  publisher={Elsevier}
}

@article{cobelli1980parameter,
  title={Parameter and structural identifiability concepts and ambiguities: a critical review and analysis},
  author={Cobelli, Claudio and Distefano 3rd, Joseph J},
  journal={American Journal of Physiology-Regulatory, Integrative and Comparative Physiology},
  volume={239},
  number={1},
  pages={R7--R24},
  year={1980},
  publisher={American Physiological Society Bethesda, MD}
}

@article{raue2009structural,
  title={Structural and practical identifiability analysis of partially observed dynamical models by exploiting the profile likelihood},
  author={Raue, Andreas and Kreutz, Clemens and Maiwald, Thomas and Bachmann, Julie and Schilling, Marcel and Klingm{\"u}ller, Ursula and Timmer, Jens},
  journal={Bioinformatics},
  volume={25},
  number={15},
  pages={1923--1929},
  year={2009},
  publisher={Oxford University Press}
}

@article{audoly2002global,
  title={Global identifiability of nonlinear models of biological systems},
  author={Audoly, Stefania and Bellu, Giuseppina and D'Angio, Leontina and Saccomani, Maria Pia and Cobelli, Claudio},
  journal={IEEE Transactions on biomedical engineering},
  volume={48},
  number={1},
  pages={55--65},
  year={2002},
  publisher={IEEE}
}

@article {durstewitz23,
    AUTHOR = {Daniel Durstewitz and Georgia Koppe and Max Ingo Thurm},
     TITLE = {Reconstructing computational system dynamics from neural data with recurrent neural networks},
   JOURNAL = {Nature Reviews Neuroscience},
      VOLUME = {24},
      YEAR = {November 2023},
     PAGES = {693--710},
}

@article {zhangtangcorpetti,
    AUTHOR = {Zheng Zhang and Ping Tang and Thomas Corpetti},
     TITLE = {Time Adaptive Optimal Transport: A Framework of Time Series Similarity Measure},
   JOURNAL = {IEEE Access},
      VOLUME = {8},
      YEAR = {2020},
     PAGES = {149764--149774},
}

@article {songyangwang16,
    AUTHOR = {H. Francis Song, Guangyu R. Yang and Xiao-Jing Wang},
     TITLE = {Training Excitatory-Inhibitory Recurrent Neural Networks for Cognitive Tasks: A Simple and Flexible Framework},
   JOURNAL = {PLoS Comput Biol},
    VOLUME = {12},
      YEAR = {2016},
    NUMBER = {2},
     URL = {http://dx.doi:10.1371/journal.pcbi.1004792},
    }

@article{tukey1977exploratory,
  title={Exploratory data analysis},
  author={Tukey, John W},
  journal={Reading/Addison-Wesley},
  year={1977}
}

@book{conover1999practical,
  title={Practical nonparametric statistics},
  author={Conover, William Jay},
  volume={350},
  year={1999},
  publisher={john wiley \& sons}
}

@article{honeycutt1992stochastic,
  title={Stochastic runge-kutta algorithms. i. white noise},
  author={Honeycutt, Rebecca L},
  journal={Physical Review A},
  volume={45},
  number={2},
  pages={600},
  year={1992},
  publisher={APS}
}

@article{ozalp2023reconstruction,
  title={Reconstruction, forecasting, and stability of chaotic dynamics from partial data},
  author={{\"O}zalp, Elise and Margazoglou, Georgios and Magri, Luca},
  journal={Chaos: An Interdisciplinary Journal of Nonlinear Science},
  volume={33},
  number={9},
  year={2023},
  publisher={AIP Publishing}
}

@incollection{deza2009encyclopedia,
  title={Encyclopedia of distances},
  author={Deza, Michel Marie and Deza, Elena},
  booktitle={Encyclopedia of distances},
  pages={1--583},
  year={2009},
  publisher={Springer}
}

@article{prinz2004similar,
  title={Similar network activity from disparate circuit parameters},
  author={Prinz, Astrid A and Bucher, Dirk and Marder, Eve},
  journal={Nature neuroscience},
  volume={7},
  number={12},
  pages={1345--1352},
  year={2004},
  publisher={Nature Publishing Group US New York}
}

@article{brunton2021modern,
  title={Modern Koopman theory for dynamical systems},
  author={Brunton, Steven L and Budi{\v{s}}i{\'c}, Marko and Kaiser, Eurika and Kutz, J Nathan},
  journal={arXiv preprint arXiv:2102.12086},
  year={2021}
}

@book{strogatz2024nonlinear,
  title={Nonlinear dynamics and chaos: with applications to physics, biology, chemistry, and engineering},
  author={Strogatz, Steven H},
  year={2024},
  publisher={Chapman and Hall/CRC}
}

@article{lorenz1963,
  author = {Lorenz, Edward N.},
  title = {Deterministic Nonperiodic Flow},
  journal = {Journal of the Atmospheric Sciences},
  volume = {20},
  number = {2},
  pages = {130--141},
  year = {1963},
  doi = {10.1175/1520-0469(1963)020<0130:DNF>2.0.CO;2}
}

@article{matsumoto2003chaotic,
  title={A chaotic attractor from Chua's circuit},
  author={Matsumoto, Takashi},
  journal={IEEE transactions on circuits and systems},
  volume={31},
  number={12},
  pages={1055--1058},
  year={2003},
  publisher={IEEE}
}

@article{chua1986double,
  title={The double scroll family},
  author={Chua, LEONO and Komuro, Motomasa and Matsumoto, Takashi},
  journal={IEEE transactions on circuits and systems},
  volume={33},
  number={11},
  pages={1072--1118},
  year={1986},
  publisher={IEEE}
}

@book{ermentrout2010mathematical,
  title={Mathematical foundations of neuroscience},
  author={Ermentrout, Bard and Terman, David M},
  volume={35},
  year={2010},
  publisher={Springer}
}

@incollection{sherwood2022fitzhugh,
  title={Fitzhugh--nagumo model},
  author={Sherwood, William Erik},
  booktitle={Encyclopedia of Computational Neuroscience},
  pages={1439--1449},
  year={2022},
  publisher={Springer}
}

@book{glass1988clocks,
  title={From clocks to chaos: The rhythms of life},
  author={Glass, Leon and Mackey, Michael C},
  year={1988},
  publisher={Princeton University Press}
}

@article{fitzhugh1960thresholds,
  title={Thresholds and plateaus in the Hodgkin-Huxley nerve equations},
  author={Fitzhugh, Richard},
  journal={The Journal of general physiology},
  volume={43},
  number={5},
  pages={867--896},
  year={1960},
  publisher={Rockefeller University Press}
}

@book{izhikevich2007dynamical,
  title={Dynamical systems in neuroscience},
  author={Izhikevich, Eugene M},
  year={2007},
  publisher={MIT press}
}

@article{zhong1994implementation,
  title={Implementation of Chua's circuit with a cubic nonlinearity},
  author={Zhong, Guo-Qun},
  journal={IEEE Transactions on Circuits and Systems I: Fundamental Theory and Applications},
  volume={41},
  number={12},
  pages={934--941},
  year={1994},
  publisher={IEEE}
}

@article{pivka1996lorenz,
  title={Lorenz equation and Chua’s equation},
  author={Pivka, Ladislav and Wu, Chai Wah and Huang, Anshan},
  journal={International Journal of Bifurcation and Chaos},
  volume={6},
  number={12b},
  pages={2443--2489},
  year={1996},
  publisher={World Scientific}
}

@book{suli2003introduction,
  title={An introduction to numerical analysis},
  author={S{\"u}li, Endre and Mayers, David F},
  year={2003},
  publisher={Cambridge university press}
}

@article{gammaitoni1998stochastic,
  title={Stochastic resonance},
  author={Gammaitoni, Luca and H{\"a}nggi, Peter and Jung, Peter and Marchesoni, Fabio},
  journal={Reviews of modern physics},
  volume={70},
  number={1},
  pages={223},
  year={1998},
  publisher={APS}
}

@article{dykman1998can,
  title={What can stochastic resonance do?},
  author={Dykman, MI and McClintock, Peter VE},
  journal={Nature},
  volume={391},
  number={6665},
  pages={344--344},
  year={1998},
  publisher={Nature Publishing Group UK London}
}

@article{pei1995detection,
  title={The detection threshold, noise and stochastic resonance in the Fitzhugh-Nagumo neuron model},
  author={Pei, Xing and Bachmann, Ken and Moss, Frank},
  journal={Physics Letters A},
  volume={206},
  number={1-2},
  pages={61--65},
  year={1995},
  publisher={Elsevier}
}

@article{moss1993stochastic,
  title={Stochastic Resonance in an Electronic FitzHugh-Nagumo Model a},
  author={Moss, Frank and Douglass, John K and Wilkens, Lon and Pierson, David and Pantazelou, Eleni},
  journal={Annals of the New York Academy of Sciences},
  volume={706},
  number={1},
  pages={26--41},
  year={1993},
  publisher={Wiley Online Library}
}

@book{Goodfellow-et-al-2016,
    title={Deep Learning},
    author={Ian Goodfellow and Yoshua Bengio and Aaron Courville},
    publisher={MIT Press},
    year={2016}
}

@article{stuart1958,
  title={On the non-linear mechanics of hydrodynamic stability},
  author={Stuart, J. T.},
  journal={Journal of Fluid Mechanics},
  volume={4},
  number={1},
  pages={1-21},
  year={1958},
}

@article{landau1944,
  title={On the problem of turbulence},
  author={Landau, L. D.},
  journal={Dokl. Akad. Nauk SSSR},
  volume={44},
  number={8},
  pages={339-349},
  year={1944},
}

\appendix

\section{Inadequacy of measures to evaluate a putative reconstruction}\label{inadequacy_appendix}
\begin{table}[H]
\begin{tabular}{lll}
{\bf A1} & {\bf A2} & {\bf A3} \\
\includegraphics[width=0.325\textwidth]{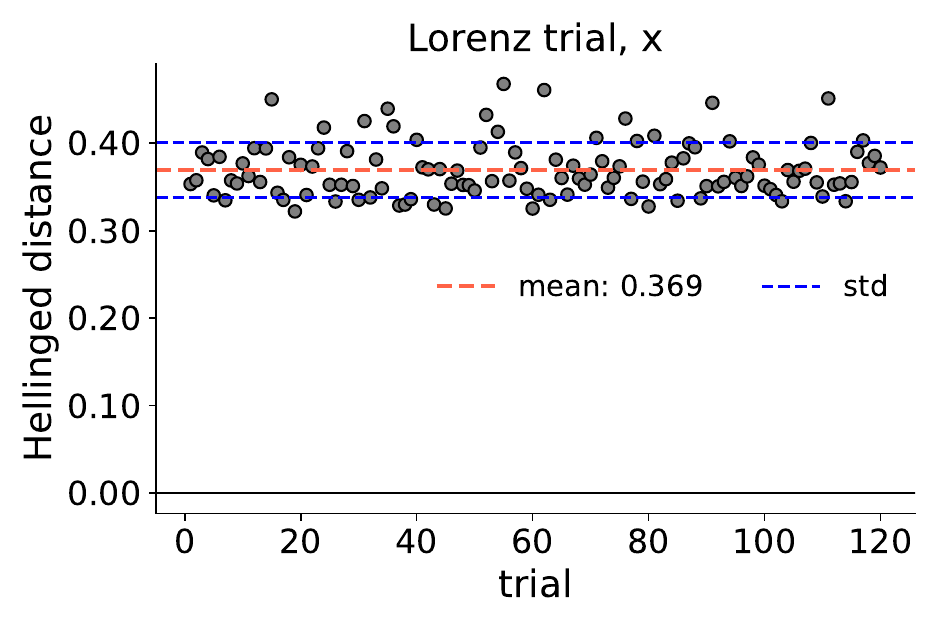} &
\includegraphics[width=0.325\textwidth]{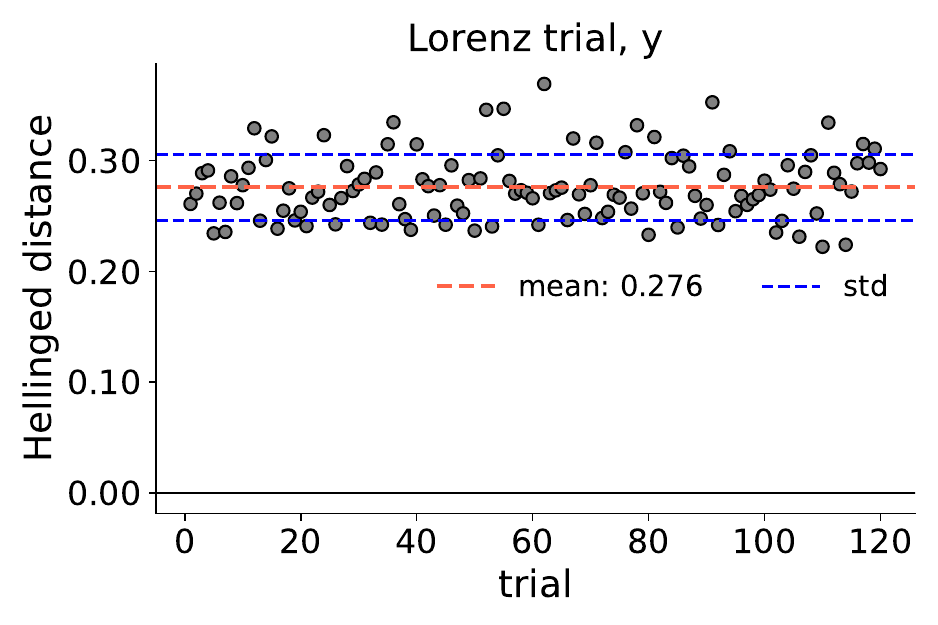}&
\includegraphics[width=0.325\textwidth]{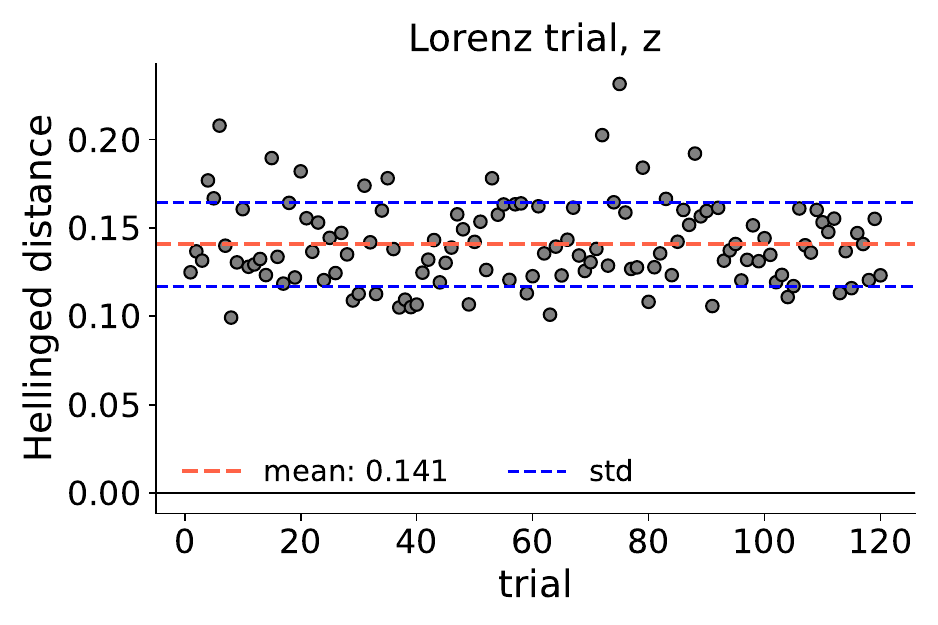} \\
{\bf B1} & {\bf B2} & {\bf B3}\\
\includegraphics[width=0.325\textwidth]{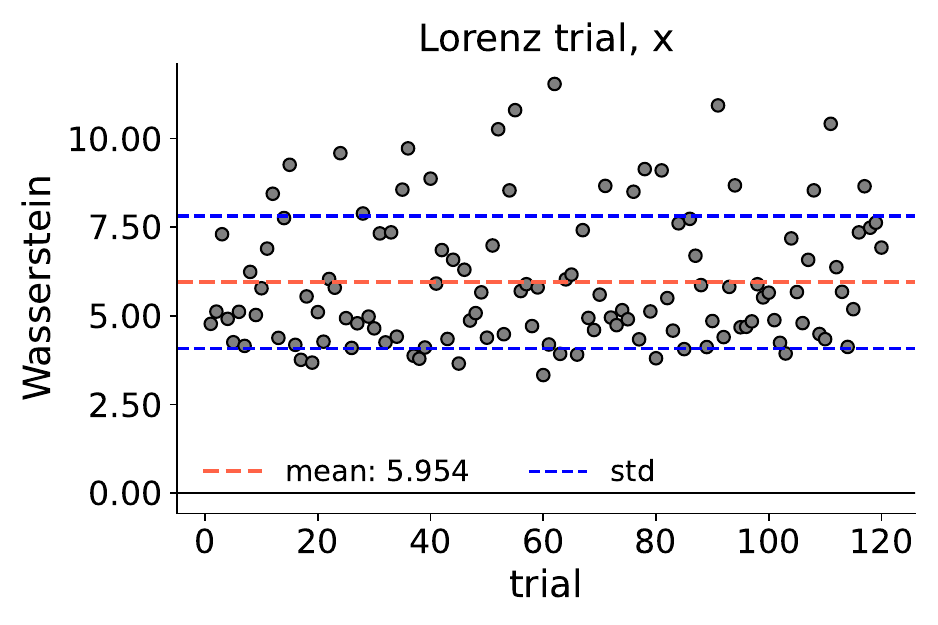} &
\includegraphics[width=0.325\textwidth]{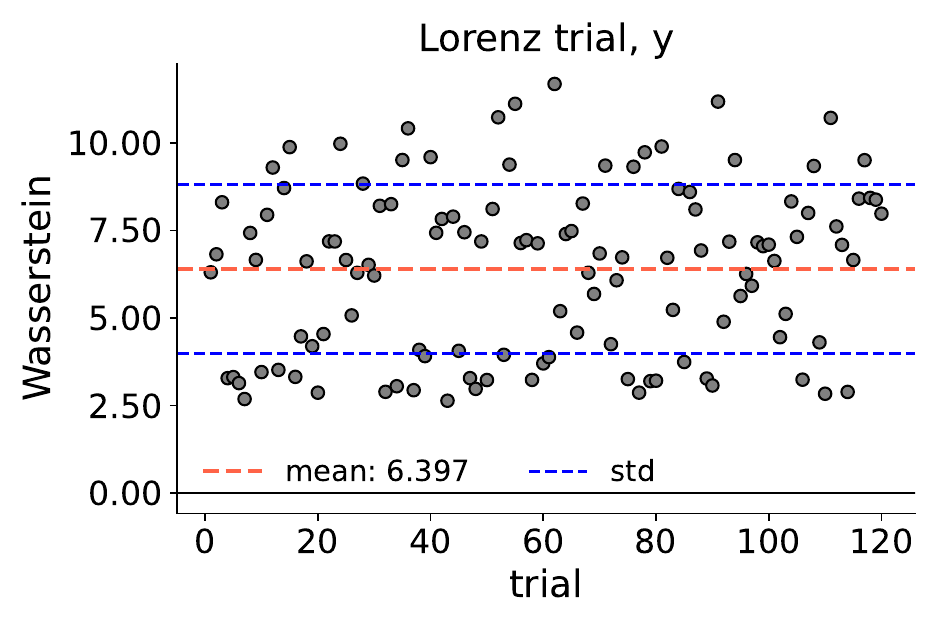}&
\includegraphics[width=0.325\textwidth]{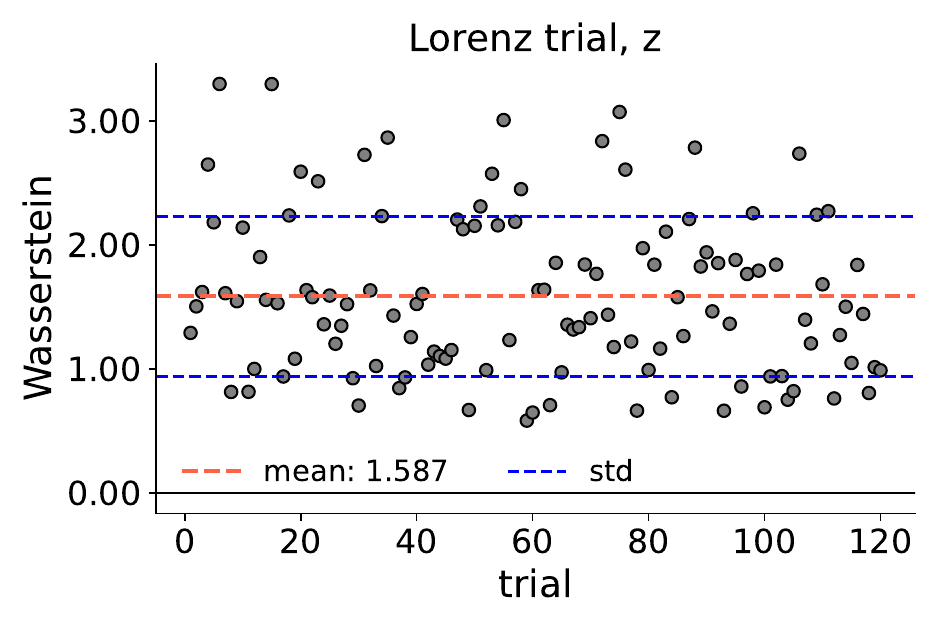} \\
\end{tabular}
\caption{\small {
{\bf A.} Hellinger distances between the histograms for the target data and trials as a function fo the trial number for the three model variables. 
{\bf B.}  Wassertein distances between the histograms for the target data and trials as a function fo the trial number for the three model variables. 
}}
\label{fig_lorenz_aux}
\end{table}

\section{Effect of noise in Lorenz and FHN system}\label{test_sta_noise}

\begin{table}[H]
\begin{tabular}{ll}
{\bf A} & \includegraphics[width=0.85\textwidth]{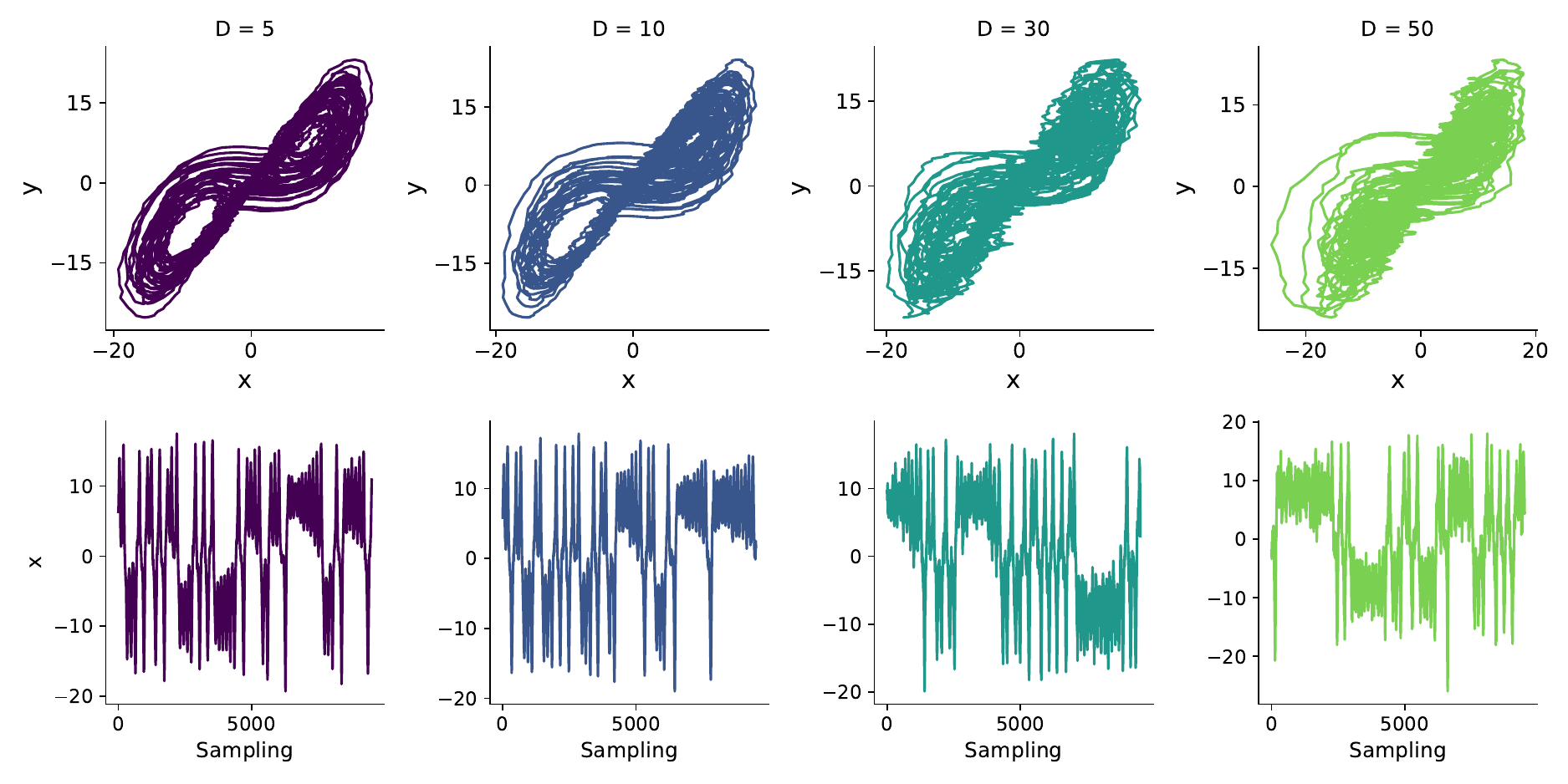}\\
{\bf B} & \includegraphics[width=0.85\textwidth]{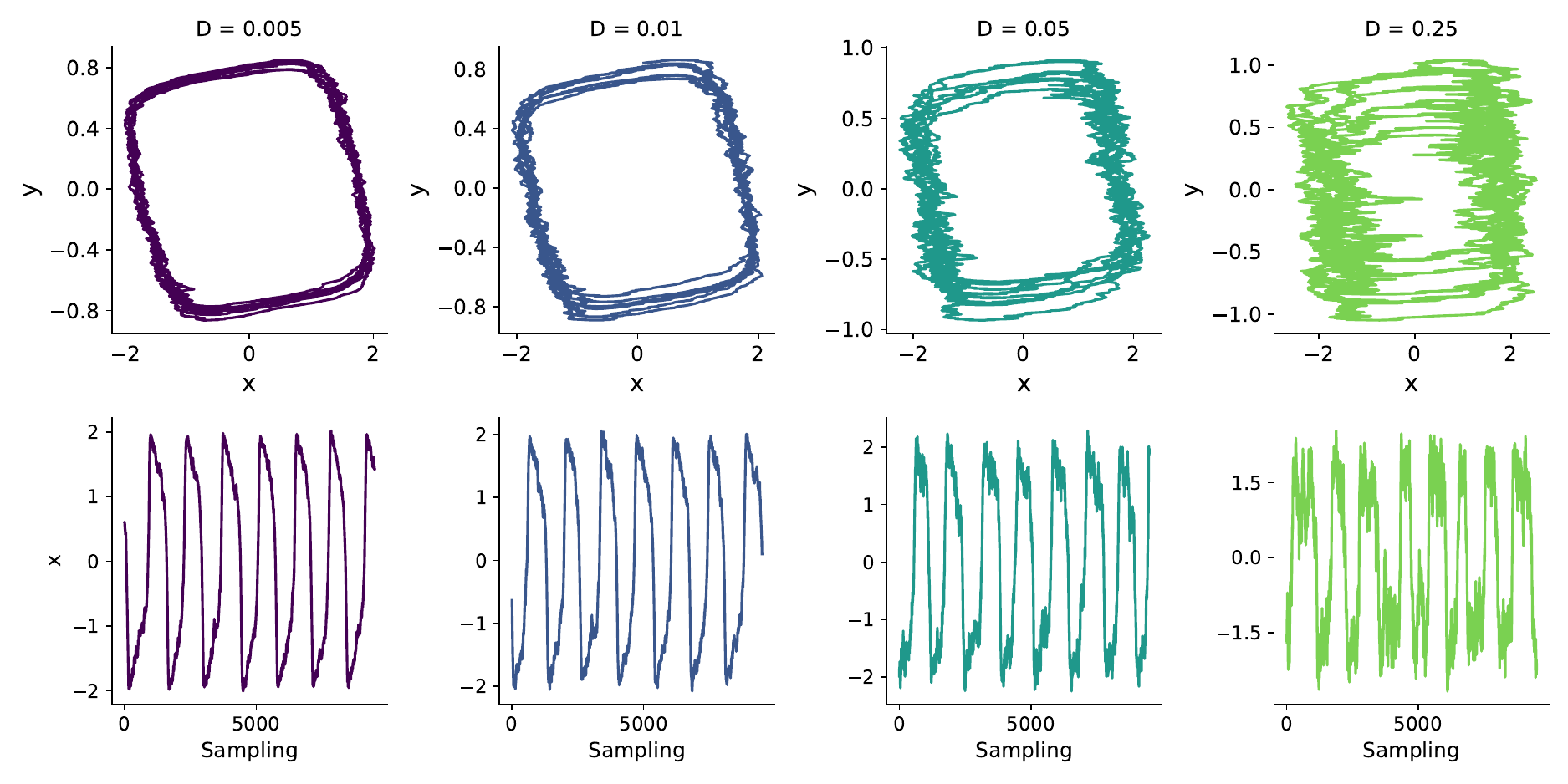} \\
\end{tabular}
\caption{{\bf Effect of noise in system used to analyze the stability of test outcome}}
\label{test_sta_orbit}
\end{table}

\section{Other result of $\Lambda\Omega$ model identification}\label{LO_others}

\begin{table}[H]
\begin{tabular}{ll}
{\bf A} & \includegraphics[width=0.5\textwidth]{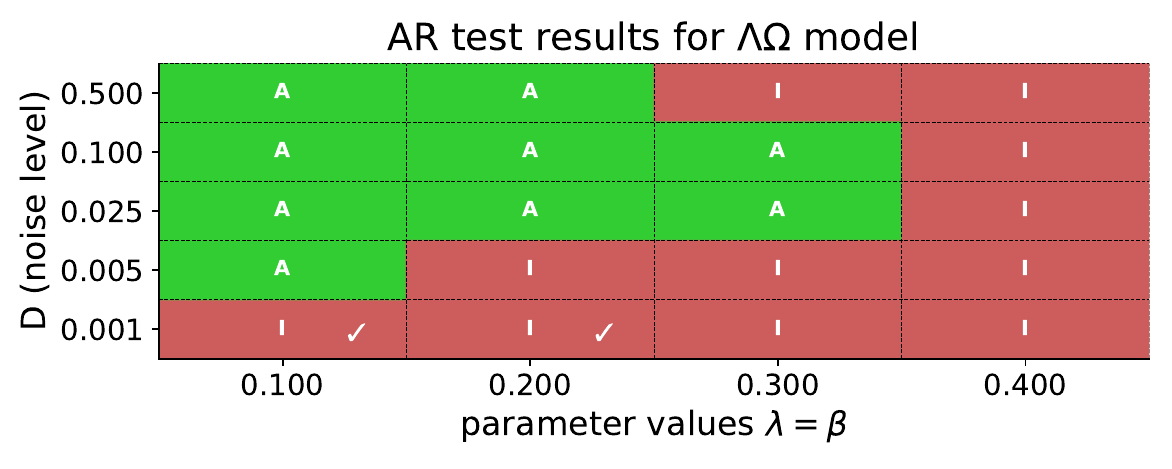} \\
{\bf B} & \includegraphics[width=0.5\textwidth]{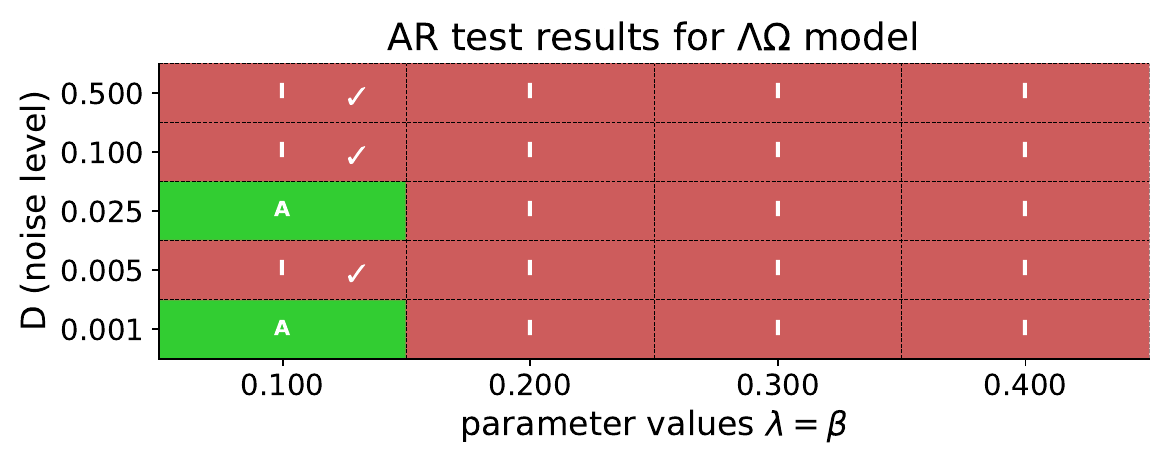} \\
{\bf C} & \includegraphics[width=0.5\textwidth]{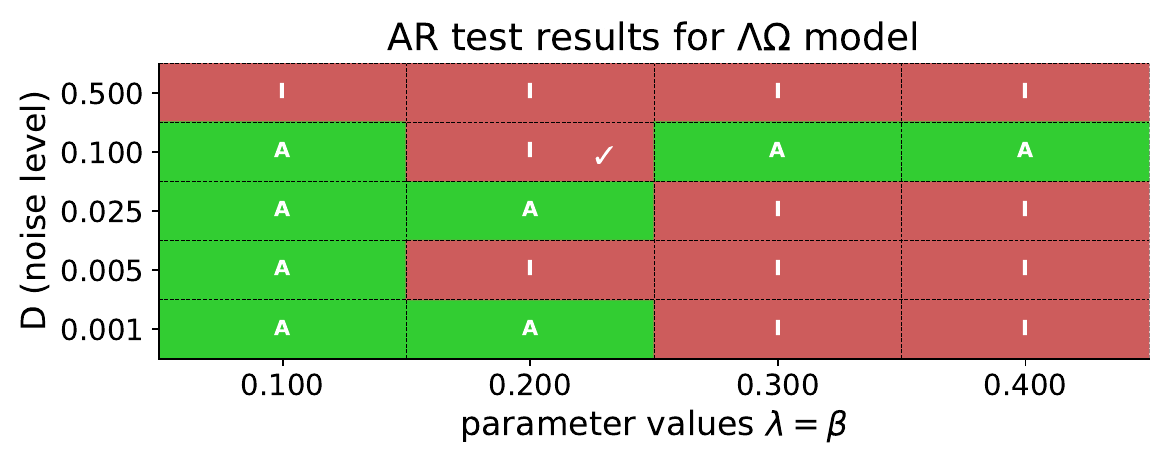} \\
\end{tabular}
\caption{{\bf $\Lambda\Omega$-model identification:} Other results obtained for $\Lambda\Omega$ systems using the same methodology as in Figure \ref{LO_system}. Structural degeneracy affects the stability of test outcome.}

\label{lorenz_deg}
\end{table}
\end{document}